\documentclass[groupedaddress,letterpaper,longbibliography,superscriptaddress,11pt]{revtex4-2}
\usepackage{graphicx}
\usepackage{tikz}
\usetikzlibrary{matrix,decorations.text,decorations.pathmorphing,decorations.markings,arrows,calc,shapes.geometric,patterns,shadows,intersections}
\usepackage{pgfplots}
\usepackage{makecell}
\usepackage{appendix}
\usepackage{amsmath,amsfonts,amssymb,mathtools}
\usepackage{hyperref}
\hypersetup{colorlinks=true,linkcolor=blue,urlcolor=blue,citecolor=blue}
\usepackage{pgfplots}
\usepackage{enumitem}

\newcommand{\marked}[1]{\textcolor{black}{#1}}

\begin{document}
\count\footins = 1000
\title{\Large{Computing hydrodynamic interactions in confined doubly-periodic geometries in linear time}}
	
\author{Aref Hashemi}
\email[]{aref@cims.nyu.edu}
\affiliation{\footnotesize{Courant Institute, New York University, New York, NY, US}}
\author{Ra\'ul P. Pel\'aez}
\affiliation{\footnotesize{Courant Institute, New York University, New York, NY, US}}
\affiliation{\footnotesize{Departamento F\'isica Te\'orica de la Materia Condensada, Universidad Aut\'onoma de Madrid, Madrid, Spain}}
\author{Sachin Natesh}
\affiliation{\footnotesize{Courant Institute, New York University, New York, NY, US}}
\affiliation{\footnotesize{Department of Applied Mathematics, University of Colorado Boulder, Boulder, CO, US}}
\author{Brennan Sprinkle}
\email[]{bsprinkl@mines.edu}
\affiliation{\footnotesize{Courant Institute, New York University, New York, NY, US}}
\affiliation{\footnotesize{Department of Applied Mathematics and Statistics, Colorado School of Mines, Golden, CO, US}}
\author{Ondrej Maxian}
\affiliation{\footnotesize{Courant Institute, New York University, New York, NY, US}}
\author{Zecheng Gan}
\affiliation{\footnotesize{Courant Institute, New York University, New York, NY, US}}
\affiliation{\footnotesize{Department of Mathematics, The Hong Kong University of Science and Technology, Hong Kong SAR, China}}
\author{Aleksandar Donev}
\email[corresponding author; ]{donev@courant.nyu.edu}
\affiliation{\footnotesize{Courant Institute, New York University, New York, NY, US}}

\begin{abstract}
We develop a linearly-scaling variant of the Force Coupling Method [K. Yeo and M. R. Maxey, J. Fluid Mech. 649, 205--231 (2010)] for computing hydrodynamic interactions among particles confined to a doubly-periodic geometry with either a single bottom wall or two walls (slit channel) in the aperiodic direction. Our spectrally-accurate Stokes solver uses the Fast Fourier Transform (FFT) in the periodic $xy$ plane and Chebyshev polynomials in the aperiodic $z$ direction normal to the wall(s). We decompose the problem into two problems. The first is a doubly-periodic subproblem in the presence of particles (source terms) with free-space boundary conditions in the $z$ direction, which we solve by borrowing ideas from a recent method for rapid evaluation of electrostatic interactions in doubly-periodic geometries [O. Maxian, R. P. Pel\'aez, L. Greengard and A. Donev, J. Chem. Phys. 154, 204107 (2021)]. The second is a correction subproblem to impose the boundary conditions on the wall(s). Instead of the traditional Gaussian kernel, we use the exponential of a semicircle kernel to model the source terms (body force) due to the presence of particles, and provide optimum values for the kernel parameters that ensure a given hydrodynamic radius with at least two digits of accuracy and rotational and translational invariance. The computation time of our solver, which is implemented in graphical processing units, scales linearly with the number of particles, and allows computations with about a million particles in less than a second for a sedimented layer of colloidal microrollers. We find that in a slit channel, a driven dense suspension of microrollers maintains the same two-layer structure as above a single wall, but moves at a substantially lower collective speed due to increased confinement.
\end{abstract}
	
\maketitle

\newcommand{\jp}[1]{{(#1)}}
\newcommand{\tp}{^\intercal}
\newcommand{\corr}{{\text{corr}}}
\newcommand{\DP}{{\text{DP}}}
\newcommand{\W}{{\text{wall}}}
\newcommand{\Set}[1]{\mathbb{#1}}
\global\long\def\Greens{\M{\Set G}}
\renewcommand{\i}{\mathsf{i}}
\global\long\def\V#1{\boldsymbol{#1}} 
\global\long\def\M#1{\boldsymbol{#1}} 
\newcommand{\sM}[1]{\M{\mathcal{#1}}} 

\global\long\def\D#1{\Delta#1} 
\global\long\def\d#1{\delta#1} 

\global\long\def\norm#1{\left\Vert #1\right\Vert }
\global\long\def\abs#1{\left|#1\right|}

\global\long\def\grad{\M{\nabla}}
\global\long\def\av#1{\left\langle #1\right\rangle}

\newcommand{\kT}{k_B T}

\global\long\def\R{\sM{R}}
\global\long\def\Rsup#1{{\sM{R}_{\mathrm{sup}}^{\mathrm{#1}}}}
\global\long\def\Rpair#1{\mathcal{R}_{\mathrm{pair},\;ij}^{\mathrm{#1}}}
\global\long\def\Rbw#1{\mathcal{R}_{\mathrm{BW},\;i}^{\mathrm{#1}}}
\global\long\def\Rsc#1{\mathcal{R}_{\mathrm{SC},\;i}^{\mathrm{#1}}}
\global\long\def\dR#1{\Delta\mathcal{R}_{\mathrm{#1},\;i}}

\global\long\def\zhat{\hat{\V z}}

\newcommand{\tabthickness}{0.5mm}
\newcolumntype{?}[1]{!{\vrule width #1}}

\section{Introduction}
The development of more efficient, accurate, and scalable methods for suspensions of rigid and flexible particles in Stokes flow remains a key challenge in soft condensed matter physics and chemical engineering. In addition to the long-ranged nature of hydrodynamic interactions, the inclusion of Brownian motion and the presence of confining boundaries pose particular difficulties. While confining boundaries partially screen the hydrodynamic interactions, they continue to decay algebraically rather than exponentially \cite{HI_Confined_Decay}, and must be captured to resolve particle dynamics.

A key component of all computational methods for Stokes flow is the efficient evaluation of the action of the singular or regularized Green's function for Stokes flow among a large number of particles. Specifically, given forces (and sometimes also torques) on many \emph{source points}, the goal is to compute the resulting (linear and sometimes also angular) velocities on many \emph{target points}, usually the same as the source points. While boundary integral methods typically use the singular Green's function, most computational methods used for large-scale suspensions are based on a regularized Green's function. There are four popular regularizations: the Rotne-Prager-Yamakawa (RPY) \cite{RPY_Yamakawa,RotnePrager} tensor, regularized Stokeslets \cite{RegularizedStokeslets}, immersed boundary (IB) kernels \cite{New6ptKernel}, and the Force Coupling Method (FCM) kernel \cite{ForceCoupling_Monopole,ForceCoupling_Stokes,ForceCoupling_Channel,ForceCoupling_Fluctuations,ForceCoupling_Lubrication}. The RPY, IB, and FCM kernels regularize (smooth) the singularity at both the source and target; this is crucial to maintain the symmetric positive definite (SPD) nature of the hydrodynamic mobility matrix, as is necessary for Brownian Dynamics (BD) methods.

Here, we develop a fast method for Stokes suspensions in \emph{doubly-periodic} (DP) geometries with one or two confining walls in the aperiodic direction. Our method is closely related to the Force Coupling Method of Yeo and Maxey \cite{ForceCoupling_Channel}, but with several important differences that increase both the flexibility and the efficiency of the approach. In brief, we employ a non-Gaussian envelope function \cite{FINUFFT_Barnett} that allows accurate computations with far fewer grid cells per particles, as in immersed boundary methods (IBM) \cite{IBM_PeskinReview}, and also develop a novel spectral Stokes solver based on recent work by some of us on fast methods for electrostatics in doubly-periodic geometries \cite{PoissonDP}.

The hydrodynamic interaction between two distinct spherical particles or \emph{blobs} of radius $a$ at positions $\V{r}^\jp{1}$ and $\V{r}^\jp{2}$ can be captured by a $3\times3$ mobility tensor that gives the velocity of one of the particles for a given force acting on the other. This hydrodynamic mobility tensor can be approximated in the far field as
\begin{equation}
\sM{O}\left(\V r^\jp{1},\V r^\jp{2}\right)=\int\delta_{a}\left(\V r^\jp{1}-\V r^{\prime}\right)\Greens\left(\V r^{\prime},\V r^{\prime\prime}\right)\delta_{a}\left(\V r^\jp{2}-\V r^{\prime\prime}\right)d\V r^{\prime}d\V r^{\prime\prime},\label{eq:RegularizedOseen}
\end{equation}
where $\Greens$ is the Green's function for Stokes flow with the specified boundary conditions. Here $\delta_{a}\left(\V r\right)$ is a regularized ``delta function'' or \emph{envelope function} \cite{ForceCoupling_Monopole} that is typically radially-symmetric, $\delta_a\left(\V{r}\right)\equiv\delta_a\left(r\right)$, when the particles are sufficiently far from boundaries. It is important to point out that the blobs do not have to represent actual physical spherical colloids; one can build physical particles as a collection of blobs, including non-spherical rigid (passive or active) particles \cite{RigidMultiblobs_Swan,SE_Multiblob_SD,StokesianDynamics_Rigid,RigidMultiblobs,BrownianMultiblobSuspensions}, which we have termed as \emph{rigid multiblobs}, or flexible particles such as semiflexible fibers \cite{FibersRPY_Keaveny,TwistBend}. If the blobs represent actual spherical colloids, additional near-field lubrication corrections can be added to improve upon the far-field accuracy \cite{BrownianDynamics_OrderNlogN,SE_Multiblob_SD,ForceCoupling_Lubrication,RollersLubrication,Broms2022}; this requires including torques in addition to forces, as we do in Sec.~\ref{sec:problem-statement}. 

Regardless of the context, a key task is to evaluate for $N$ blobs their far-field velocity $\V{U}$ from the applied forces $\V{F}$ through the action of the \emph{mobility matrix} $\sM{M}$, $\V{U}=\sM{M}\V{F}$,
\begin{equation}
\forall i:\quad \V{U}^\jp{i}=\sum_{j=1}^{N}\sM{O}\left(\V r^\jp{i},\V r^\jp{j}\right)\V F^\jp{j},\label{eq:U_from_F}
\end{equation}
in time linear in the number of blobs $N$. Note that the form of \eqref{eq:RegularizedOseen} guarantees that the mobility matrix is SPD by construction since $\Greens$ is an SPD kernel, and the regularization is applied both at the source and the target.

The specific choice of $\delta_a$ as a delta function on the surface of a sphere of radius $a$, $\delta_a\left(\V{r}\right)=\left(4\pi a^2\right)^{-1}\delta\left(r-a\right)$, results in the widely-used Rotne-Prager-Yamakawa (RPY) tensor \cite{StokesianDynamics_Wall,RPY_Shear_Wall,SpectralRPY}. When the kernels do not overlap the boundaries, one can transform \eqref{eq:RegularizedOseen} into a differential form by employing the Fax\'en differential operator,
\begin{equation}
\sM{O}\left(\V{r}^\jp{1},\V{r}^\jp{2}\right)\approx\left(\M{I}+\frac{a^2}{6}\grad_{\V{r}^{\prime}}^2\right)\left(\M{I}+\frac{a^2}{6}\grad_{\V{r}^{\prime\prime}}^2\right)\Greens(\V{r}^{\prime},\V{r}^{\prime\prime})\big|_{\V{r}^{\prime\prime}=\V{r}^\jp{2}}^{\V{r}^{\prime}=\V{r}^\jp{1}}.\label{eq:MobilityFaxen}
\end{equation}
This analytical simplification allows for explicit evaluation of the RPY kernel, not just in an unbounded domain but also in a half-space above a no-slip wall \cite{StokesianDynamics_Wall}, because for a single no-slip boundary there is a relatively simple image construction for $\Greens$ due to Blake \cite{blake1971note}. This image construction has also enabled the development of fast methods for evaluating \eqref{eq:U_from_F} for a single wall, based on either (flexible periodicity) Fast Multipole Method \cite{STKFMM,FMM_wall} or the Fast Fourier Transform (FFT) \cite{SpectralEwald_Wall,SE_MixedPeriodicityStokes}. For triply-periodic (TP) domains, the Positively Split Ewald (PSE) method \cite{SpectralRPY} for evaluating \eqref{eq:U_from_F} (and also generating Brownian velocities) for the RPY kernel provides the basis for Fast Stokesian Dynamics \cite{SE_Multiblob_SD}. However, it remains a challenge to construct a similarly-efficient method for confined suspensions since PSE uses Fourier representations in a key way for each component of the method (Stokes solver, generating Brownian increments, and Ewald splitting).

In principle, a greater flexibility with respect to boundary conditions can be achieved by replacing analytical Green's functions with a grid-based Stokes solver \cite{BrownianBlobs}. This requires replacing the singular surface delta function form of $\delta_a\left(\V{r}\right)$ with a smooth function that can be resolved on a grid. In the FCM, a Gaussian envelope function $\delta_a$ is used, which allows for analytical calculation of $\sM{O}$ in an unbounded domain \cite{ForceCoupling_Monopole,ForceCoupling_Stokes}; in numerical methods the Gaussian is truncated. By contrast, in the IBM, $\delta_a$ is a discrete grid function specifically constructed to maximize grid invariance \cite{IBM_PeskinReview,New6ptKernel,RigidMultiblobs}, and the double convolution in \eqref{eq:RegularizedOseen} is discrete. This makes all IBM results grid- and solver-dependent, without a direct continuum limit.

We develop a method that combines favorable features of FCM and IBM. Namely, we maintain the continuum representation \eqref{eq:RegularizedOseen} from FCM, however, we do not use a Gaussian envelope but rather use the ``exponential of a semi-circle'' (ES) kernel proposed by Barnett for its efficiency in the context of non-uniform FFTs \cite{FINUFFT_Barnett}. Not only does this kernel allow for greater flexibility in tuning the hydrodynamic radius of the particles \cite{DISCOS_Periodic}, but also allows us to use many fewer grid cells per particle than for a Gaussian kernel, comparable to the IBM, while still solving the continuum equations to several digits of accuracy. Specifically, for about two-three digits of accuracy we need four grid cells per particle (in each dimension) if only translational velocity is required, and five or six if rotational velocities are also required. The grid independence of the results makes them transferable, and allows us to separate the Stokes solver from the physics (i.e., the problem has a solution without a discretization). We find that $\sM{O}$ computed with the ES and Gaussian kernels are indistinguishable in practice. We employ an image construction following Yeo and Maxey \cite{ForceCoupling_Channel} (also used in the IBM \cite{RigidIBM,RigidMultiblobs}) to generalize \eqref{eq:RegularizedOseen} to the relevant-in-practice case when some of the blob kernels overlap the boundaries; this is considerably harder to do for the RPY kernel, and has required \emph{ad hoc} fixes in past work by some of us \cite{MagneticRollers}.

For TP domains one can easily solve the Stokes equations spectrally in Fourier space, but this is considerably harder when boundaries are present. In Sec.~\ref{sec:stokes-problem} we use some of the ideas applied to the Poisson equation in \cite{PoissonDP} to develop a Stokes solver for DP geometries that are unbounded or confined by one or two boundaries in the aperiodic direction. Specifically, we focus on \emph{bottom wall} (BW) DP geometries, with a single wall at $z=0$, and \emph{slit channel} (SC) DP geometry, confined by two walls at $z=0$ and $z=H$, but the Stokes solver can handle more flexible boundary conditions in the aperiodic direction. Our novel fluid solver makes it possible to handle geometries that are partially unbounded in one direction, unlike in existing IBM or FCM implementations based on more traditional grid solvers like finite differences or finite elements \cite{ForceCoupling_SpectralSolver}. Our fluid solver has the additional advantage that its implementation requires only calls to the three-dimensional (3D) FFT with an oversampling factor of $2$ in the aperiodic direction, combined with trivially parallelizable one-dimensional pentadiagonal boundary value solvers in the $z$ direction. We use this to implement the method on Graphical Processing Units, achieving linear scaling up to as many as one million particles, with scaling constants much better than existing fast methods \cite{FMM_wall} for the types of problems we study here. It is important to note that Srinivasan and Tornberg have recently developed a sophisticated method based on 3D FFTs for the singular Stokes Green's function with a bottom wall geometry (using images) and \emph{flexible} periodicity \cite{SpectralEwald_Wall,SE_MixedPeriodicityStokes}. Our approach is different and specialized to doubly-periodic geometries, is conceptually simpler, and affords flexibility in the boundary conditions in the unbounded direction. It is beyond the scope of this work to give a thorough comparison of the different approaches.

In Sec.~\ref{sec:stokes-problem} we also study how to choose the parameters of the ES kernel to achieve a desired accuracy and effective blob hydrodynamic radius with the coarsest possible solver grid. In Sec.~\ref{sec:valid} we perform a number of validation tests examining the self and pair mobility of particles in BW and SC geometries, and show that the method produces results in agreement with existing theoretical or numerical predictions. In Sec.~\ref{sec:colloidallayers} we demonstrate that it is possible to generate stochastic (Brownian) particle displacements with covariance proportional to $\sM{M}$ using a Lanczos method \cite{SquareRootKrylov,SquareRootPreconditioning} in a modest number of iterations independent of the number of particles, for both BW and SC geometries. This allows us to replace the core hydrodynamic routines used in the Stokesian dynamics method developed in \cite{RollersLubrication} and the rigid multiblob BD methods developed in \cite{BrownianMultiblobSuspensions} with new linear-scaling implementations that are substantially more efficient for sufficiently large number of particles.

These new computational developments enable us to perform larger-scale studies of the dynamics of confined microroller suspensions than previously feasible, in Sec.~\ref{sec:colloidallayers}. After validating that the results presented by some of us in \cite{RollersLubrication} are free of finite-size effects, we explore what a top wall does to a driven dense suspension of colloidal microrollers. We find that while the suspension maintains a two-layer structure as above a single bottom wall, in a slit channel the collective velocity is substantially reduced due to the increased confinement, as expected.

\section{Model Formulation}\label{sec:problem-statement}
We develop a method to solve the Stokes equations,
\begin{align}
  \eta\grad^2\V{u}-\grad p & =-\V{f},\label{eq:momentum}\\
  \grad\cdot\V u & = 0,\label{eq:continuity}
\end{align}
in an $xy$ doubly-periodic domain of size $\begin{bmatrix}x & y\end{bmatrix}\in[-L_x,L_x]\times[-L_y,L_y]$ \footnote{For our results, we consider a square domain in $xy$ plane for simplicity, i.e., $L_{x/y}=L$.}, with $z\in[0,\infty)$; we refer to this as a \emph{bottom wall} (BW) geometry. We will return later to the case of a \emph{slit channel} (SC) geometry with two walls at $z=0$ and $z=H$, for which $z\in[0,H]$. Here $\V{u}=\begin{bmatrix}u & v & w\end{bmatrix}\tp$ is the velocity field of the fluid, $p$ the pressure, $\eta$ the viscosity, and the body force $\V{f}=\begin{bmatrix}f & g & h\end{bmatrix}\tp$ represents particles. To close the problem, we impose tangential slip boundary conditions (BCs) at $z=0$, i.e.,
  \begin{equation}
    \V{u}|_{z=0}=\V{u}^\W=\begin{bmatrix}u^\W & v^\W & 0\end{bmatrix}\tp,\label{eq:bcnoslip0}
  \end{equation}
  where $\V{u}^\W(x,y)$ is a smooth function giving a prescribed slip velocity ($\V{u}^\W=\V{0}$ for no-slip boundaries) along the wall (e.g., electrophoretic slip), and we require boundedness of $\V{u}$ as $z\to\infty$. For all of the results and tests presented in this study, we use $\V{u}^\W=\V{0}$.

  The source term $\V{f}$ is restricted to the domain limits (i.e. $\V{f}(z\notin[0,H])=\V{0}$) and consists of regularized monopoles and dipoles with envelope functions $\Delta_{M}$ and $\Delta_{D}$ centered around particles or blobs, i.e.,
  \begin{equation}
    \V{f}(\V{x})=\sum_{j=1}^{N}\left[\V{F}^\jp{j}\Delta_{M}(\V{x}-\V{r}^\jp{j})+\frac{1}{2}\grad\times(\V{\tau}^\jp{j}\Delta_{D}(\V{x}-\V{r}^\jp{j}))\right],\label{eq:fcm}
  \end{equation}
  where $\V{x}=\begin{bmatrix}x & y & z\end{bmatrix}\tp$ and $\V{r}^\jp{j}=\begin{bmatrix}x^\jp{j} & y^\jp{j} & z^\jp{j}\end{bmatrix}\tp$ is the $j$th particle's location with $j=1,\dots,N$, and $\V{F}^\jp{j}$ and $\V{\tau}^\jp{j}$ are the force and torque of the $j$th particle. Here, $\Delta_M$ and $\Delta_D$ are compactly-supported, smooth, regularized delta functions called kernels in the immersed boundary method (IBM), or envelopes in the Fast Coupling Method (FCM). In the classical FCM, the envelopes are (truncated) Gaussians,
  \begin{equation}
    \Delta_{M/D}(\V{x}) =\frac{1}{\sqrt{8\pi^{3}{g_{M/D}}^{6}}}\exp\left(-\frac{\norm{\V{x}}^{2}}{2g_{M/D}^{2}}\right),\label{eq:kernmd}
  \end{equation}
  where $g_M=R_h/\sqrt{\pi}$ and $g_D=R_h/\left(6\sqrt{\pi}\right)^{\tfrac{1}{3}}$, with $R_h$ as the effective hydrodynamic radius of a particle/blob.
  
  If a particle is closer to the wall than a distance $z_{\mathrm{im}}$ so that its kernel overlaps the wall, we add the negative of the particle's envelope centered at its image point about the wall. This ensures that the force/torque decays to zero as a particle approaches the wall. As a result, if $z^\jp{j}<z_{\mathrm{im}}$, we replace the envelopes $\Delta_{M}$ and $\Delta_{D}$ in \eqref{eq:fcm} with
  \begin{equation}
    \Delta^{W}_{M/D}\left(\V{x}-\V{r}^\jp{j}\right)=\Delta_{M/D}\left(\V{x}-\V{r}^\jp{j}\right)-\Delta_{M/D}\left(\V{x}-\V{r}_{\text{im}}^\jp{j}\right),\label{eq:kernelnearwall}
  \end{equation}
  where $\V{r}_{\text{im}}^\jp{j}=\V{r}^\jp{j}-2\hat{\V{e}}_z(\hat{\V{e}}_z\cdot\V{r}^\jp{j})$ is the particle's point of reflection about the bottom wall, and $\hat{\V{e}}_z=\begin{bmatrix}0 & 0 & 1\end{bmatrix}\tp$. In the FCM, Yeo and Maxey \cite{ForceCoupling_Channel} put a positive sign in \eqref{eq:kernelnearwall} for $\Delta_D$, to ensure that the angular velocity of a particle in simple shear flow is unaffected by the proximity to the wall. We put a negative sign to ensure that all components of the particle mobility go to zero at the wall. This is a modeling choice, and it is important to realize that \eqref{eq:kernelnearwall} is not an exact image construction for a no-slip wall \cite{SpectralEwald_Wall}. For a free-slip boundary such as a gas-liquid interface, it is straightforward to make an exact image construction; see, for example, Appendix A in \cite{FluctuatingFCM_DC}.

  Our goal is to compute the motion of each of the particles for dynamical simulations. Suppose we've solved the Stokes equations for the fluid velocity $\V{u}$; then the linear and angular velocities of the particles, denoted as $\V{U}^\jp{j}$ and $\V{\Omega}^\jp{j}$, can be obtained through the following volume integrals over the envelopes,
  \begin{align}
    \V{U}^\jp{j} & =\int_{\V{x}}\V{u}(\V{x})\Delta_M(\V{x}-\V{r}^\jp{j})d\V{x},\label{eq:getpartvel}\\
    \V{\Omega}^\jp{j} & =\frac{1}{2}\int_{\V{x}}\left(\grad\times\V{u}(\V{x})\right)\Delta_D(\V{x}-\V{r}^\jp{j})d\V{x},\label{eq:getpartvort}
  \end{align}
  with the understanding that $\Delta_{M/D}$ get replaced by $\Delta_{M/D}^W$ for particles close to a wall.

  In our own variant of FCM, we will not use a Gaussian kernel; rather, we will employ a compactly-supported Exponential of a Semicircle (ES) kernel \cite{FINUFFT_Barnett} optimized for numerical efficiency in spectral methods, as we discuss in detail in Sec.~\ref{sec:es}. Other kernels can be used as well, as long as their Fourier transform decays sufficiently fast in the wavenumber.
  
  \section{Stokes Problem}\label{sec:stokes-problem}
  We develop a doubly-periodic+correction Stokes solver based on the observation that, the original problem \eqref{eq:momentum}--\eqref{eq:continuity} can be separated into two subproblems, namely the doubly-periodic ``DP'' problem and the ``correction'' problem. For the DP subproblem, we keep the source terms $\V{f}$ but there is no wall present (i.e., open BCs in $z$). The correction subproblem, has the walls but has no forcing, so it can be solved analytically using a plane-wave expansion. The total solution is then the sum of the solutions to the two subproblems.

  Below, we detail our solution method for the bottom wall (BW) geometry with a no-slip wall. Appendix~\ref{sec:appendixA-sc} summarizes the method for the slit channel (SC) geometry with somewhat more general partial slip BCs. The Stokes solver can also handle open boundaries in the $z$ direction, if the total force on the domain is zero. In order to uniquely define the self mobility of a particle, one must therefore add a negative tail to the kernel to account for the \emph{backflow} around the particle, similar to how in TP domains the $\V{k}=\begin{bmatrix}k_x & k_y & k_z\end{bmatrix}\tp=\V{0}$ component of the solution is set to zero, where $\V{k}$ is the wavenumber. We defer discussion of backflow in unbounded DP geometries to future work.

  \subsection{Doubly-periodic Stokes solver}\label{sec:dpsolver}
  For the ``DP'' subproblem, we solve a doubly-periodic Stokes problem for $\V{u}_{\text{DP}}$ and $p_{\text{DP}}$:
  \begin{align}
    \eta\grad^{2}\V{u}_{\text{DP}}-\grad p_{\text{DP}} & =-\V{f},\label{eq:momentum-dp}\\
    \grad\cdot\V{u}_{\text{DP}} & =0.\label{eq:continuity-dp}
  \end{align}
  Here the source term $\V{f}$ is the same as \eqref{eq:momentum}, and the doubly-periodic domain is $\begin{bmatrix}x & y & z\end{bmatrix}\in[-L_x,L_x]\times[-L_y,L_y]\times(-\infty,+\infty)$ with \emph{free-space} BCs in $z$, i.e., $p_\DP$ and $\V{u}_{\text{DP}}$ are bounded as $z\to\pm\infty$ and there is no wall present. Note that in our spectral DP solver, we will solve $\V{u}_{\text{DP}}$ in a bounded domain $\begin{bmatrix}x & y & z\end{bmatrix}\in[-L_x,L_x]\times[-L_y,L_y]\times[0,H]$ for either bottom wall or slit channel geometry. For a slit channel, \eqref{eq:kernelnearwall} ensures that $\V{f}$ is smooth on and vanishes outside of $z\in[0,H]$, and for bottom wall case, we will assume that it is possible to find a maximum height $H$, so that $\V{f}$ smoothly goes to zero for $z\notin[0,H]$. Also, note that the system represented by \eqref{eq:momentum-dp} and \eqref{eq:continuity-dp} is not well-posed if $\int_{\V{x}}\V{f}(\V{x})d\V{x}\neq \V{0}$, where the integration is over the simulation box $[-L_x,L_x]\times[-L_y,L_y]\times[0,H]$; however, this problem manifests in our method only for $\begin{bmatrix}k_x & k_y\end{bmatrix}\tp=\V{0}$, which we will handle separately in Sec.~\ref{sec:k0mode}. As we now show, the open boundary condition can be reduced to $z\in[0,H]$ through the Dirichlet-to-Neumann map, as proposed for the Poisson equation by Maxian et al.~\cite{PoissonDP}.

  \subsubsection{Solution for $p_\DP$}
  First take the divergence of \eqref{eq:momentum-dp} and use \eqref{eq:continuity-dp} to obtain
\begin{equation}
  \grad^2p_\DP=\grad\cdot\V{f}.\label{eq:poisson}
\end{equation}
Fourier transforming in the $xy$ domain yields 
\begin{equation}
  \left(\partial_z^2-k^2\right)\hat{p}_\DP(\V{k},z)=\begin{bmatrix}\i\V{k} \\ \partial_z\end{bmatrix}\cdot\hat{\V{f}}(\V{k},z).\label{eq:poissonfs}
\end{equation}
Here $\V{k}=\begin{bmatrix}k_x & k_y\end{bmatrix}\tp=\begin{bmatrix} n_x & n_y\end{bmatrix}\tp\pi/L_{x/y}$ with $n_{x/y}\in\Set{Z}$, and $k=\norm{\V{k}}$. For $z\not\in[0,H]$, $\V{f}=\V{0}$ and the general solution to \eqref{eq:poissonfs} is
\begin{equation}
  \hat{p}_\DP(\V{k},z\ge H)=C_1e^{-k z},\quad\hat{p}_\DP(\V{k},z\le 0)=C_2e^{k z},\label{eq:psolfs-outside}
\end{equation}
where we used the boundedness of $\hat{p}_\DP$ as $z\to\pm\infty$. The solutions \eqref{eq:psolfs-outside} imply simple Dirichlet-to-Neumann maps for $\hat{p}_\DP$ at $z=0^-$ and $z=H^+$. Since $\hat{p}_\DP$ is continuously differentiable across the (artificial) computational boundaries at $z=0$ and $z=H$, the same maps apply at $z=H^-$ and $z=0^+$ as well, giving
\begin{equation}
  \left(\partial_z+k\right)\hat{p}_\DP(\V{k},z=H)=0,\quad \left(\partial_z-k\right)\hat{p}_\DP(\V{k},z=0)=0.\label{eq:DtoNmapp}
\end{equation}
Therefore, we can solve the boundary value problem (BVP) \eqref{eq:poissonfs} on $z\in[0,H]$ subject to the boundary conditions given by \eqref{eq:DtoNmapp}. This approach was proposed by some of us for the Poisson equation in \cite{PoissonDP}.

\subsubsection{Solution for $\V{u}_\DP$}
To find the velocity, we take an $xy$ Fourier transform of \eqref{eq:momentum-dp}:
\begin{equation}
  -\eta\left(\partial_z^2-k^2\right)\hat{\V{u}}_\DP(\V{k},z)+\begin{bmatrix}\i\V{k} \\ \partial_z\end{bmatrix}\hat{p}_\DP(\V{k},z)=\hat{\V{f}}(\V{k},z).\label{eq:momentumfs}
\end{equation}
Again, $\V{f}=\V{0}$ outside domain $z\in[0,H]$, giving
\begin{align}
  \partial_z^2\hat{u}_\DP-k^2\hat{u}_\DP&=\frac{\i k_x}{\eta}\hat{p}_\DP,\label{eq:momentumfs-outside-u}\\
  \partial_z^2\hat{w}_\DP-k^2\hat{w}_\DP&=\frac{1}{\eta}\partial_z\hat{p}_\DP.\label{eq:momentumfs-outside-w}
\end{align}
Using the solution for $\hat{p}_\DP$ in \eqref{eq:psolfs-outside}, and enforcing boundedness of $\hat{u}_\DP$ as $z\to\pm\infty$, we find the general solution for $\V{k}\neq\V{0}$,
\begin{subequations}
\label{eq:usolfs-outside}
\begin{align}
    \hat{u}_\DP(\V{k},z\ge H)&=A_xe^{-kz}-\i\frac{k_xC_1}{2k\eta}ze^{-kz},\\
    \hat{u}_\DP(\V{k},z\le 0)&=B_xe^{kz}+\i\frac{k_xC_2}{2k\eta}ze^{kz}.
\end{align}
\end{subequations}
Due to $xy$ symmetry of the problem, solutions to $\hat{v}_\DP$ are analogous and obtained by replacing $x$ with $y$ in \eqref{eq:usolfs-outside}.

From \eqref{eq:usolfs-outside}, we get the following Dirichlet-to-Neumann maps at the boundaries:
\begin{equation}
  \left(\partial_z+k\right)\begin{bmatrix}\hat{u}_\DP(\V{k},H)\\\hat{v}_\DP(\V{k},H)\end{bmatrix}=-\frac{\i\V{k}}{2k\eta}\hat{p}_\DP(\V{k},H),\quad\left(\partial_z-k\right)\begin{bmatrix}\hat{u}_\DP(\V{k},0)\\\hat{v}_\DP(\V{k},0)\end{bmatrix}=\frac{\i\V{k}}{2k\eta}\hat{p}_\DP(\V{k},0).\label{eq:DtoNmapuv}
\end{equation}

Similarly, the general solution for $\hat{w}_\DP$ is
\begin{subequations}
\label{eq:wsolfs-outside}
\begin{align}
    \hat{w}_\DP(\V{k},z\ge H)&=A_ze^{-kz}+\frac{C_1}{2\eta}ze^{-kz},\\
    \hat{w}_\DP(\V{k},z\le 0)&=B_ze^{kz}+\frac{C_2}{2\eta}ze^{kz}.
\end{align}
\end{subequations}
which implies the following boundary conditions:
\begin{equation}
  \left(\partial_z+k\right)\hat{w}_\DP(\V{k},H)=\frac{1}{2\eta}\hat{p}_\DP(\V{k},H),\quad\left(\partial_z-k\right)\hat{w}_\DP(\V{k},0)=\frac{1}{2\eta}\hat{p}_\DP(\V{k},0).\label{eq:DtoNmapw}
\end{equation}

We can solve the BVP problems \eqref{eq:momentumfs} on $z\in[0,H]$ with boundary conditions \eqref{eq:DtoNmapuv} and \eqref{eq:DtoNmapw} to find the velocity field using a spectral (Chebyshev) solver, as described in Sec.~\ref{sec:stokes-solver}. It is important to note that solving the pressure and velocity BVPs only requires solving \emph{pentadiagonal} linear systems.

  \subsection{Correction solve}\label{sec:correctionsolve}
  The correction subproblem solves the following homogeneous Stokes problem for $\V{u}_\corr$ and $p_\corr$,
  \begin{align}
    \eta\grad^{2}\V{u}_\corr-\grad p_\corr & =0,\label{eq:momentum-corr}\\
    \grad\cdot\V{u}_\corr & =0,\label{eq:continuity-corr}
  \end{align}
  where the doubly-periodic domain is $\begin{bmatrix}x & y & z\end{bmatrix}\in[-L_x,L_x]\times[-L_y,L_y]\times[0,+\infty)$ and with slip on the bottom wall
  \begin{equation}
    \V{u}_\corr|_{z=0}=\V{u}^\W-\V{u}_\DP|_{z=0},\label{eq:bcnoslip0-corr}
  \end{equation}
  where $\V{u}_\DP$ is the solution to the DP problem. Note that, by linearity, it is clear that the sum of the solution to the two subproblems in the upper-half plane gives the solution to the original problem \eqref{eq:momentum}--\eqref{eq:bcnoslip0}, i.e.,
  \begin{align}
    \V{u}=\V{u}_\DP+\V{u}_\corr,\quad\text{and}\quad p=p_\DP+p_\corr.\label{eq:superposition}
  \end{align}
  The correction subproblem \eqref{eq:momentum-corr}--\eqref{eq:bcnoslip0-corr} can be solved analytically.

\subsubsection{Solution for $p_\corr$}
Taking the divergence of \eqref{eq:momentum-corr} we obtain a Laplace equation for pressure
\begin{equation}
  \Delta p_\corr=0.\label{eq:laplace}
\end{equation}
After a Fourier transform in $xy$, we get a 1D equation for each Fourier mode $\V{k}$,
\begin{equation}
  \partial_z^2\hat{p}_\corr(\V k,z)-k^{2}\hat{p}_\corr(\V k,z)=0.\label{eq:laplacefs}
\end{equation}
Using the boundedness of $\hat{p}_\corr$ as $z\rightarrow\infty$, the general solution of \eqref{eq:laplacefs} is
\begin{equation}
  \hat{p}_\corr(\V{k},z)=C_{0}e^{-kz}.\label{eq:psolfs-corr}
\end{equation}

\subsubsection{Solution for $\V{u}_\corr$}
Taking the Fourier transform of \eqref{eq:momentum-corr} in $xy$ and using \eqref{eq:psolfs-corr} we obtain
\begin{align}
  \eta\left(k^{2}-\partial_z^2\right)\hat{u}_\corr(\V{k},z) & =-\i k_{x}C_0e^{-kz},\\
  \eta\left(k^{2}-\partial_z^2\right)\hat{w}_\corr(\V{k},z) & =kC_0e^{-kz},
\end{align}
with general solutions for $\V{k}\ne\V{0}$ given by
\begin{align}
  \hat{u}_\corr(\V{k},z) & =-\frac{C_{0}\i k_{x}}{2\eta k}ze^{-kz}+C_xe^{-kz},\label{eq:usolfs-corr}\\
  \hat{w}_\corr(\V{k},z) & =\frac{C_{0}}{2\eta}ze^{-kz}+C_ze^{-kz},\label{eq:wsolfs-corr}
\end{align}
and analogously for $\hat{v}_\corr$. To determine the unknown coefficients, we consider the boundary conditions on the wall. Let $\V{u}_\corr|_{z=0}=\V{u}^\W-\V{u}_\DP|_{z=0}=\begin{bmatrix}u_0(x,y) & v_0(x,y) & w_0(x,y)\end{bmatrix}\tp$, i.e.,
\begin{equation}
  \hat{u}_\corr(\V{k},0)=\hat{u}_0(\V{k}),\quad\hat{v}_\corr(\V{k},0)=\hat{v}_0(\V{k}),\quad\hat{w}_\corr(\V{k},0)=\hat{w}_0(\V{k}).\label{eq:bcnoslip0-corr-scalarfs}
\end{equation}
Substituting \eqref{eq:usolfs-corr}--\eqref{eq:wsolfs-corr} into \eqref{eq:bcnoslip0-corr-scalarfs} gives three of the unknown coefficients:
\begin{equation}
C_x=\hat{u}_0(\V{k}),\quad C_y=\hat{v}_0(\V{k}),\quad C_z=\hat{w}_0(\V{k}).
\end{equation}

To find the coefficient $C_0$ we take an $xy$ Fourier transform of \eqref{eq:continuity-corr}: 
\begin{equation}
\i k_{x}\hat{u}_\corr(\V{k},z)+\i k_{y}\hat{v}_\corr(\V{k},z)+\partial_{z}\hat{w}_\corr(\V k,z)=0.\label{eq:continuityfs-corr}
\end{equation}
Evaluating \eqref{eq:continuityfs-corr} at $z=0$ and also using the boundary conditions \eqref{eq:bcnoslip0-corr-scalarfs}, we obtain 
\begin{equation}
\partial_{z}\hat{w}_\corr(\V{k},z=0) =-\i k_{x}\hat{u}_0(\V{k})-\i k_{y}\hat{v}_0(\V{k}),
\end{equation}
which coupled with \eqref{eq:wsolfs-corr} gives
\begin{equation}
  C_0=2\eta(k\hat{w}_0(\V{k})-\i k_{x}\hat{u}_0(\V{k})-\i k_{y}\hat{v}_0(\V{k})).
\end{equation}

Substituting the coefficients back to \eqref{eq:usolfs-corr}--\eqref{eq:wsolfs-corr}, we obtain an explicit correction solution in $xy$ Fourier space:
\begin{align}
  \hat{p}_\corr(\V{k},z) & =2\eta(k\hat{w}_0(\V{k})-\i k_{x}\hat{u}_0(\V{k})-\i k_{y}\hat{v}_0(\V{k}))e^{-kz},\label{eq:psolfs-corr-final}\\  
  \hat{u}_\corr(\V{k},z) & =-\frac{k_x}{k}(\i k\hat{w}_0(\V{k})+k_{x}\hat{u}_0(\V{k})+k_{y}\hat{v}_0(\V{k}))ze^{-kz}+\hat{u}_0(\V{k})e^{-kz},\label{eq:usolfs-corr-final}\\
  \hat{v}_\corr(\V{k},z) & =-\frac{k_y}{k}(\i k\hat{w}_0(\V{k})+k_{x}\hat{u}_0(\V{k})+k_{y}\hat{v}_0(\V{k}))ze^{-kz}+\hat{v}_0(\V{k})e^{-kz},\label{eq:vsolfs-corr-final}\\
  \hat{w}_\corr(\V{k},z) & =(k\hat{w}_0(\V{k})-\i k_{x}\hat{u}_0(\V{k})-\i k_{y}\hat{v}_0(\V{k}))ze^{-kz}+\hat{w}_0(\V{k})e^{-kz}.\label{eq:wsolfs-corr-final}
\end{align}

\subsection{Zero mode ($\V{k}=\V{0}$)}\label{sec:k0mode}
Sections~\ref{sec:dpsolver} and \ref{sec:correctionsolve} provide the DP and correction solutions for $\V{k}\neq \V{0}$. We treat the $\V{k}=\V{0}$ mode here for the overall solution without separating it into subproblems.

  To find $\hat{w}(\V{0},z)$, we use the continuity equation \eqref{eq:continuity} in Fourier space for $k_x=k_y=0$,
  \begin{equation}
    \partial_z\hat{w}(\V{0},z)=0.
  \end{equation}
  As a result, $\hat{w}(\V{0},z)$ is just a constant, and the no-slip boundary conditions at $z=0$, given in Fourier space by
  \begin{equation}
    \hat{\V{u}}(\V{k},0)=\begin{bmatrix}\hat{u}^\W(\V{k}) & \hat{v}^\W(\V{k}) & 0\end{bmatrix}\tp,\label{eq:bcnoslip0fs}
  \end{equation}
  indicate that
  \begin{equation}
    \hat{w}(\V{0},z)=0.
    \label{eq:wsolfs-k0}
  \end{equation}
  The momentum equation \eqref{eq:momentum} in Fourier space for $\V{k}=\V{0}$ can be expressed as
  \begin{equation}
    -\eta\partial_z^2\hat{\V{u}}(\V{0},z)+\begin{bmatrix}\V{0} \\ \partial_z\end{bmatrix}\hat{p}(\V{0},z)=\hat{\V{f}}(\V{0},z)=\begin{bmatrix}\hat{f}(\V{0},z) \\ \hat{g}(\V{0},z) \\ \hat{h}(\V{0},z)\end{bmatrix}.\label{eq:momentumfs-k0}
  \end{equation}
  Using \eqref{eq:wsolfs-k0}, the $z$ component of \eqref{eq:momentumfs-k0} simplifies to
  \begin{equation}
    \partial_z\hat{p}(\V{k}=0,z)=\hat{h}(\V{k}=0,z),
  \end{equation}
  which yields the zero mode solution for the pressure,
  \begin{equation}
    \hat{p}(\V{0},z)=\int_{0}^{z}\hat{h}(\V{0},z^{\prime})dz^{\prime},
    \label{eq:psolfs-k0}
  \end{equation}
  where we set $\hat{p}(\V{0},0)=0$ as a convention.

  To find the $\V{k}=\V{0}$ mode solution for $\hat{u}$ (or $\hat{v}$ due to symmetry in $xy$), note that the $x$ component of \eqref{eq:momentumfs-k0} is
  \begin{equation}
    \partial_z^2\hat{u}(\V{0},z)=-\frac{\hat{f}(\V{0},z)}{\eta}.\label{eq:momentumfs-k0-u}
  \end{equation}
  We have a no-slip BC at $z=0$ given by \eqref{eq:bcnoslip0fs}. To find a boundary condition at $z=H$, note that for $z>H$, $\partial_{z}^2\hat{u}(\V{0},z)=0$, which along with the boundedness of velocity as $z\to\infty$, indicate that $\hat{u}(\V{0},z\geq H)$ is some constant. Since $\partial_z\hat{u}(\V{0},z)$ is continuous across $z=H$ and it is a constant for $z>H$, we conclude that $\partial_{z}\hat{u}(\V{0},H)=0$. As a result, the $\V{k}=\V{0}$ mode of $\hat{u}$ can be computed by solving the BVP \eqref{eq:momentumfs-k0-u} with the following boundary conditions:
  \begin{equation}
    \hat{u}(\V{0},0)=\hat{u}^\W(\V{0}),\quad \partial_z\hat{u}(\V{0},H)=0,
    \label{eq:bvp-u-k0-bcs}
  \end{equation}
  and analogously for $\hat{v}$.

\subsection{Stokes solver}\label{sec:stokes-solver}
Our spectral doubly-periodic Stokes solver is based on the method proposed for the Poisson equation in \cite{PoissonDP}, where more details are provided. We use FFT in the doubly-periodic $xy$ plane and Chebyshev polynomials in the $z$ direction to achieve spectral accuracy. The grid is uniform in the $xy$ plane, with spacing $h_{x/y}=2L_{x/y}/N_{x/y}$ where $N_{x/y}$ is the total number of points in each direction, and nonuniform in the $z$ direction, with grid points placed on a type-2 Chebyshev grid including the endpoints at $z=0$ and $z=H$. The number of Chebyshev grid points, $N_z$, is chosen such that the coarsest spacing in the $z$ direction (occurring at the midplane) is comparable to $h_{x/y}$ (see Eq.~(108) in \cite{PoissonDP}).

In principle, we could take a 2D FFT of \eqref{eq:momentum-dp} and \eqref{eq:poisson} to find the discrete forms of \eqref{eq:momentumfs} and \eqref{eq:poissonfs}, respectively, and solve the resulting BVPs in the $z$ direction in Chebyshev space. Instead, following Maxian et al.~\cite{PoissonDP}, we take advantage of the Chebyshev compatibility with the Fourier transform \footnote{Chebyshev coefficients of a function can be obtained by evaluating its values at the Chebyshev points and taking a Fast Cosine Fourier Transform \cite{FastCosineTransform,GreengardBVP}. Our implementation uses a 3D FFT with $2N_z-2$ points along the $z$ direction for simplicity. An alternative approach is introduced in \cite{FastCosineTransform} that requires no upsampling at all at the cost of some increased complexity; such a fast DCT is, however, not implemented in the cuFFT GPU library (but is in the FFTW CPU library).} and solve the problem in 3D Fourier-Chebyshev space \cite{SpectralMATLAB}. This allows us to use 3D FFTs in the implementation, which gives greater flexibility for optimizing parallel performance. We have implemented the method in CUDA for Graphical Processing Units (GPUs) in the UAMMD library \cite{UAMMD}, using the NVIDIA cuFFT library. Both CPU and GPU implementations are freely available at \url{https://github.com/stochasticHydroTools/DoublyPeriodicStokes}.

The following steps summarize our solution algorithm for the Stokes equation:
\begin{enumerate}[label=(\arabic*),ref=\#\arabic*]
\item\label{solverItem:spread} Spread the right hand side (RHS) $\V{f}$ of the momentum equation \eqref{eq:momentum-dp} on the grid using the ES kernel (cf. Sec.~\ref{sec:es}). Since the ES kernel is compactly supported, each particle only spreads to $\sim m^3$ grid points. In the $z$ direction this is only true if near the midplane; more than $m$ points will be required near the boundaries where the Chebyshev grid is finer. When the kernel extends outside of the domain over a physical boundary \footnote{Kernels must never extend outside open boundaries, which can always be placed at an appropriate $z$ location to ensure this.}, we fold the contribution outside of the domain back into the domain using a negative mirror image (see \eqref{eq:kernelnearwall}). We will return to spreading torques after this list.

\item Take a 3D FFT of $\V{f}$ to compute the Chebyshev coefficients of $\hat{\V{f}}(\V{k},z)$. Then, use Chebyshev differentiation to compute the Chebyshev coefficients of $\partial_z\hat{\V{f}}(\V{k},z)$ \cite{SpectralMATLAB}.

\item For each $\V{k}\neq\V{0}$, solve the BVP \eqref{eq:poissonfs} with boundary conditions \eqref{eq:DtoNmapp} for the Chebyshev coefficients of $\hat{p}_\DP(\V{k},z)$. Then, evaluate $\hat{p}(\V{k},z=0/H)$ by direct summation to assemble boundary conditions \eqref{eq:DtoNmapuv} and \eqref{eq:DtoNmapw}, and solve the BVP \eqref{eq:momentumfs} for the Chebyshev coefficients of $\hat{\V{u}}_\DP(\V{k},z)$. See Appendix A in \cite{PoissonDP} for details of the BVP solver, which is based on \cite{GreengardBVP}; this efficient and stable solver only requires solving pentadiagonal systems.

\item Evaluate, by direct summation, $\hat{\V{u}}_\DP(\V{k},z=0)$ for the BW geometry and also $\hat{\V{u}}_\DP(\V{k},z=H)$ for the SC geometry \footnote{For partial slip walls, also compute $\partial_z\hat{\V{u}}_\DP(\V{k},z=0/H)$ by Chebyshev differentiation and direct summation.}, and compute the RHSs of \eqref{eq:bcnoslip0-corr-scalarfs} and \eqref{eq:bcslip-corr}, respectively. Evaluate the correction solutions, $\hat{p}_\corr(\V{k},z)$ and $\hat{\V{u}}_\corr(\V{k},z)$ on the Chebyshev grid, using the analytical expressions \eqref{eq:psolfs-corr-final}--\eqref{eq:wsolfs-corr-final} for BW and \eqref{eq:psolfs-corr-sc}--\eqref{eq:wsolfs-corr-sc} for SC geometries.

\item Evaluate the $\V{k}=\V{0}$ mode solution for $\hat{p}(\V{0},z)$ using \eqref{eq:psolfs-k0} and Chebyshev integration, and solve the BVP given by \eqref{eq:momentumfs-k0-u} and \eqref{eq:bvp-u-k0-bcs} for the BW and \eqref{eq:bvp-uv-k0-sc} for the SC geometry to find $\hat{u}(\V{0},z)$, and analogously, $\hat{v}(\V{0},z)$. Set $\hat{w}(\V{0},z)=0$.
  
\item For all $\V{k}$, take a 1D FFT in the $z$ direction to compute the Chebyshev coefficients of the correction and zero-mode solutions and add them to $\hat{p}_\DP$ and $\hat{\V{u}}_\DP$.

\item Back transform to real space using a 3D iFFT to get $p(\V{x})$ and $\V{u}(\V{x})$.

\item Interpolate $\V{u}(\V{x})$ at the particle positions via \eqref{eq:getpartvel} and trapezoidal+Chebyshev quadrature to find the linear velocity of the particles. For each particle, this requires the same grid values (but with additional $z$-dependent quadrature weights) as in the spreading in step \ref{solverItem:spread}.
\end{enumerate}

When torques are also applied, they can be spread in two ways. The first way, which is more straightforward to implement but also more computationally expensive, is to compute the derivatives needed for the curl operator in Fourier-Chebyshev space; this requires an additional forward 3D FFT. Analogously, in the interpolation step, the curl of the velocity is first computed in Fourier-Chebyshev space, and then transferred back to real space with an additional 3D iFFT. These additional FFTs can be avoided by instead using the derivative of the ES kernel to spread and interpolate velocity \cite{SE_MixedPeriodicityStokes} \footnote{Since the derivative of the ES kernel is unbounded near the boundaries of its support, we have implemented this by truncating the derivative to zero for arguments with magnitude larger than the point where the derivative has minimal magnitude before it blows up. See \cite{SE_MixedPeriodicityStokes} for a more sophisticated treatment.}. We have employed both of these approaches to spread torques and interpolate velocity in a CPU-based implementation and confirm that both preserve the desired accuracy. Due to the higher implementation complexity of the second approach \footnote{In particular, it requires using a different kernel along different dimensions. Additionally, to ensure all components of the particle mobility go to zero at a wall, we must use a positive sign for the mirror image for the $z$ partial derivative of the ES kernel.}, in the GPU code we have presently only implemented the first way.

It is important to emphasize that all steps in the method scale linearly, or log-linearly, in either the number of particles or the number of grid points. The BVP solutions along the $z$ direction require solving pentadiagonal linear systems (see Appendix A in \cite{PoissonDP}), which makes that step need only a linear amount of memory and computational effort in $N_z$. Combined with the independence of the BVP solves for each $\V{k}$, this makes the algorithm particularly well-suited for efficient execution on GPUs. 

\subsection{Exponential of a semicircle kernel (ES)}\label{sec:es}
In order to reduce the costs associated with representing particles on the grid, we replace the Gaussian kernel traditionally used in the FCM formulation with the exponential of a semicircle (ES) kernel \cite{FINUFFT_Barnett} \footnote{See also \cite{SE_MixedPeriodicityStokes} for a related use of an ES-like kernel as a window function in spectral Ewald methods.}. Specifically, as in the IBM, we take $\Delta_{M/D}$ to be tensor products of a scalar kernel $\phi$:
\begin{equation}
  \label{eq:es}
  \Delta_{M/D}\left(\V{x};\alpha_{M/D},\beta_{M/D}\right)=\prod_{i=1}^3\phi\left(x_i;\alpha_{M/D},\beta_{M/D}\right),
\end{equation}
where
\begin{equation}
  \phi(z;\alpha,\beta)=\left(\int_{-\alpha}^{\alpha}e^{\beta\left(\sqrt{1-\left(\frac{z}{\alpha}\right)^{2}}-1\right)}dz\right)^{-1}
  \begin{cases}
    e^{\beta\left(\sqrt{1-\left(\frac{z}{\alpha}\right)^{2}}-1\right)},\quad & \abs{\frac{z}{\alpha}}\leq1\\
    0,\quad & \text{otherwise}.
  \end{cases}
  \label{eq:es-scalar}
\end{equation}
Here, $(\alpha,\beta)$ are fixed parameters uniquely tied to the kernel, which is compactly supported on $[-\alpha,\alpha]$. If the kernel we use for spreading and interpolation has a simple Fourier transform, we can determine the effective hydrodynamic radius $R_h$ of a particle represented by such a kernel analytically. For example, in the seminal works on FCM applied to Stokes flow \cite{ForceCoupling_Monopole,ForceCoupling_Stokes}, the authors show that for Gaussian monopole and dipole envelopes given by \eqref{eq:fcm} with standard deviations $g_M$ and $g_D$, the length scale of the kernels is related to the particle radius through $g_M=R_{h}/\sqrt{\pi}$ and $g_D=R_{h}/(6\sqrt{\pi})^{1/3}$. However, the ES kernel does not have a simple analytical Fourier transform, so we will numerically determine the relationship between $R_{h}$ and $(\alpha,\beta)$.

Let us denote with $m$ the number of grid cells in the support of the kernel in each dimension. For a regular grid with spacing $h$, $\alpha\le\tfrac{1}{2}hm$ for the ES kernel, but we typically take $\alpha=\tfrac{1}{2}hm$. Kernels like the Peskin kernels used in IBM \cite{New6ptKernel} have no free parameters to change $R_h$ once $m$ is chosen. In contrast, the ES kernel affords increased flexibility with the additional parameter $\beta$. So, particles of different radii can be represented with the same support $\alpha$, but different $\beta$ values. Moreover, the ES kernel has numerically compact support (to some tolerance) in real and Fourier space, and is, therefore, well suited for spectral discretizations of the Stokes problem. That is, the discrete RHS of the Stokes problem \eqref{eq:momentum} will decay rapidly in Fourier space.

To find an appropriate value for $\beta$, we consider the mobility of an isolated sphere. A sphere of radius $R_h$ will translate and rotate with linear and angular velocities $\V{U}$ and $\V{\Omega}$ in an \emph{unbounded} space under the action of a force $\V{F}=6\pi\eta R_h\V{U}$ and torque $\V{\tau}=8\pi\eta R_h^3\V{\Omega}$. For a TP domain with a cubic box of size $L^3$, periodic corrections to Stokes' law can be expressed as an asymptotic expansion in $R_h/L$ \cite{Mobility2D_Hasimoto},
\begin{equation}
  \label{eq:tp-velocity-linear-angular}
\V{F}\approx\frac{6\pi\eta R_h\V{U}}{1-K_M\frac{R_{h}}{L}},\quad\V{\tau}\approx\frac{8\pi\eta R_h^{3}\V{\Omega}}{1-K_D\left(\frac{R_{h}}{L}\right)^3},
\end{equation}
to leading order, where, $K_M$ and $K_D$ are constants independent of $R_h$ and $L$. Our numerical results indicate that the same form applies to doubly-periodic domains as well, but with different constants; in this section we use a TP system so that the grid spacing $h$ is uniform.

\begin{figure}[t]
  \centering
  \includegraphics{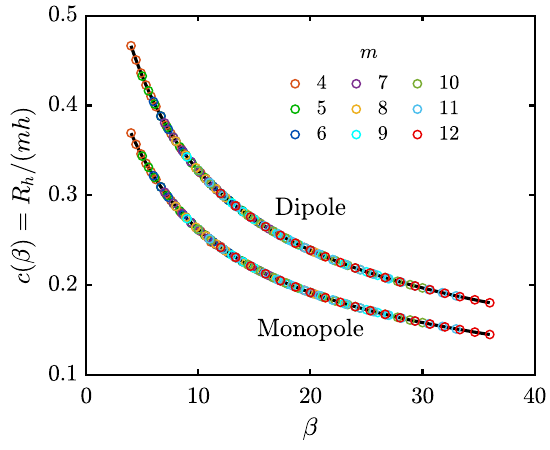}
  \caption{Dimensionless hydrodynamic radius function $c(\beta)$ for the ES kernel for translation and rotation. For $30$ particle positions distributed uniformly in one grid cell, and for several values of $m$ and $\beta$, we extrapolate the computed radii to $L\to\infty$. Specifically, we take $h=1$, $L\in\{60,95,130,165,200\}$, $m\in\{4,5,6,7,8,9,10,11,12\}$, and $\beta/m$ values linearly spaced in $[1,3]$. All of the data collapse on a master curve (depicted in solid black) that can be fit by a polynomial, up to a small error due to loss of translational invariance.}
  \label{fig:cbetafit}
\end{figure}

By the scale invariance of the ES kernel under changing the unit of length there exists a function $c(\beta)$ such that
\begin{equation}
  R_h(\alpha,\beta)=2\alpha\;c(\beta)=(hm)\;c(\beta).
  \label{eq:radform}
\end{equation}
To obtain $c(\beta)$ empirically, and determine an ``optimal'' $\beta$, we take a single particle in a TP domain and apply a unit force or torque on it, and measure its linear and angular velocities. If we keep the position of the particle (relative to the grid) and $\alpha$ fixed, we can use extrapolation to $L\to\infty$ based on \eqref{eq:tp-velocity-linear-angular} to obtain $R_h$. Due to numerical errors, $R_h$ varies slightly based on the exact position of the particle relative to the grid, resulting in numerical loss of translational invariance. In Fig.~\ref{fig:cbetafit} we show $c(\beta)$ estimated from the mean $\av{R_h}$, along with a polynomial fit. We emphasize that $c(\beta)$ is an intrinsic property of the ES kernel, and is independent of the Stokes solver used, as long as the solver is sufficiently accurate. This is assured to about three digits by our use of a spectral solver for any $m\ge4$; smaller $m$ will have too few points per particle to resolve the envelope. The loss of translational invariance is purely numerical and gets worse for smaller $m$.

\begin{figure}[t]
  \centering
  \includegraphics{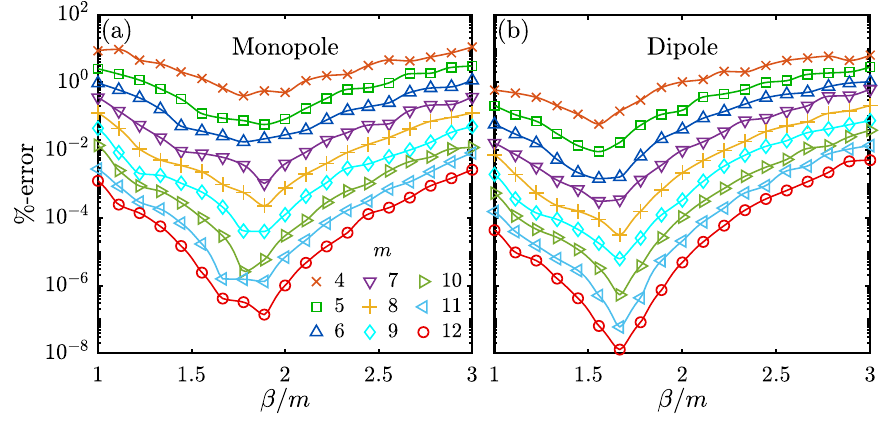}
  \caption{Percent error (variability) in the effective hydrodynamic radius of a particle represented by the ES kernel as the particle moves relative to an underlying triply-periodic (TP) uniform grid of spacing $h$. For a given width $m$, there is an optimal $\beta/m$ which yields the smallest spread in the calculated effective radius (caused by the loss of translational invariance).}
  \label{fig:effRad_err}
\end{figure}

In Fig.~\ref{fig:effRad_err}, we evaluate percent errors in the extrapolated radii at each $m$ and for each $\beta$ in terms of a $95\%$ confidence interval. That is, we report
\begin{equation}
\label{eq:percenterror}
\%\;\text{error}=\frac{4\sigma(m)}{\av{R_h(m)}}\times100\%,
\end{equation}
in which $\sigma(m)$ is the standard deviation from the mean radius $\av{R_{h}(m)}$. We see that for the monopole, $m=12$ and $\beta/m\approx1.9$ gives the smallest \%-error, while for the dipole, $m=12$ and $\beta/m\approx1.7$ are the optimal settings. From an extrapolation with these highly resolved settings, we find $K_M=2.83$ and $K_D=4.19$ \footnote{The value of $K_m$ compares well to the first term in the higher order periodic correction of Hasimoto \cite{Mobility2D_Hasimoto}. The value of $K_D$ is also very close to $4\pi/3$ as predicted in \cite{zuzovsky1983spatially} but with an opposite sign, which we believe to be a misprint in \cite{zuzovsky1983spatially}.}.

Note, in practice and for the sake of efficiency, we use $m=4,5,6$, for which $\beta/m\approx1.75$ and $\beta/m\approx1.6$ are optimal settings for the monopole and dipole kernels, respectively. This is to be compared to the optimal value $\beta/m\approx2.7$ suggested by Barnett et al.~\cite{FINUFFT_Barnett} for the non-uniform FFT algorithm. The precise values of $\beta$ and $m$ for each particle species need to be selected by balancing several considerations. First, decreasing $R_h/h=c(\beta)m$ means fewer grid cells in the support of the particle kernel, and therefore, higher efficiency. However, translational invariance ought to be preserved to some target tolerance, requiring keeping $\beta/m$ in a narrow range around the minima of error seen in Fig.~\ref{fig:effRad_err}. Lastly, $L/h$ must be an integer, preferably an FFT-friendly integer. The target tolerance for us is $2$--$3$ digits of accuracy. A final challenge is that the hydrodynamic radius needs to be matched between the monopole and dipole terms. If we use $m=4$ for the monopole, which is the smallest possible to ensure sufficient grid invariance, we find that we require $m\geq5$ for the dipole to match $R_h$ and get acceptable accuracy. Table~\ref{tab:optes} summarizes values we suggest for $m=5$ and $m=6$ (it is best to keep $m$ the same for the monopole and dipole kernels). When only translational motion is required (e.g., with the rigid multiblob method \cite{RigidMultiblobs}), $m=4$ can be used; Table~\ref{tab:optes_trans} summarizes the suggested values. Our public domain software release contains a script that suggests ``good'' values for $m$ and $\beta$ given $L$ and $R_h$ as inputs. When there are more than one particle species, the selection becomes more involved but still possible \cite{DISCOS_Periodic}.

\begin{table}[h]
\parbox[t]{.45\linewidth}{
\caption{Optimal combinations of $m_M,m_D,\beta_M,\beta_D$ along with minimal errors in translational invariance \eqref{eq:percenterror} of the effective hydrodynamic radius $R_h$ (equal for monopole \& dipole up to $3$ digits), when both force and torque are applied ($M\;\&\;D$).}
\begin{center}
  \begin{tabular}{c?{\tabthickness}c|c}
    $m_M(=\!m_D)$ & $5$ &  $6$\\
    \Xhline{\tabthickness}
    $R_h/h$              & $1.560$ & $1.731$\\
    $\beta_M/m_M$        & $1.305$ & $1.327$\\
    $\beta_D/m_D$        & $2.232$ & $2.216$\\
    $\%\;\text{error}_M$ & $0.90$ & $0.15$\\
    $\%\;\text{error}_D$ & $0.81$ & $0.21$
  \end{tabular}
\end{center}
\label{tab:optes}
}
\hfill
\parbox[t]{.45\linewidth}{
\caption{Same as Table~\ref{tab:optes} but only forces are applied ($M$).}
\begin{center}
  \begin{tabular}{c?{\tabthickness}c|c|c}
    $m_M$ & $4$ &  $5$ & $6$\\
    \Xhline{\tabthickness}
    $R_h/h$              & $1.205$ & $1.344$ & $1.554$\\
    $\beta_M/m_M$        & $1.785$ & $1.886$ & $1.714$\\
    $\%\;\text{error}_M$ & $0.37$ & $0.05$ & $0.02$
  \end{tabular}
\end{center}
\label{tab:optes_trans}
}
\end{table}

\section{Self and Pair Mobilities}\label{sec:valid}
We validate our solver against theoretical predictions on basic mobility problems in the bottom wall (BW) and slit channel (SC) geometries with $N=1$ or $N=2$ particles. Specifically, for several particle configurations, we determine the mobility matrix $\sM{M}\in\Set{R}^{6N\times6N}$, relating velocities to forces,
\begin{equation}
  \begin{bmatrix}\V{U}\\\V{\Omega}\end{bmatrix}
  =\begin{bmatrix}\sM{M}^{tt} & \sM{M}^{tr}\\\sM{M}^{rt} & \sM{M}^{rr}\end{bmatrix}
  \begin{bmatrix}\V{F}\\\V{\tau}
  \end{bmatrix},
  \label{eq:grandmob}
\end{equation}
where $\sM{M}^{(..)}\in\Set{R}^{3N\times3N}$ with superscripts $t$ and $r$ denoting the translational and rotational mobility components, respectively, and $\V{F},\V{\tau}\in\Set{R}^{3N}$ and $\V{U},\V{\Omega}\in\Set{R}^{3N}$ are the vectors of force and torque on the particles and the resulting translation and rotational velocities. We will refer to certain elements of $\sM{M}$ using the notation $\mu_{ab}^{cd}$ for diagonal blocks (self mobility) or $\nu_{ab}^{cd}$ for off-diagonal blocks (pair mobility). This corresponds to prescribing a force or torque ($d=t\text{ or }r$) in the $b$ direction $(b=x,y,\text{ or }z)$ on one particle, and measuring the linear or rotational velocity ($c=t\text{ or }r$) in the $a$ direction on the second particle (or the same particle, in the case of self mobility). For example, $\nu_{yx}^{tr}$ corresponds to the linear velocity in the $y$ direction of the second particle due to a torque about the $x$ axis on the first particle, i.e., $U_y^\jp{2}=\nu_{yx}^{tr}\tau_x^\jp{1}$. Note that $\nu_{ab}^{cd}=\mu_{ab}^{cd}$ if the two particles have the same position.

In all tests, we use a box with dimensions $L_{x/y}=L=76.8R_h,\;H=19.2R_h$ to allow direct comparison with prior IBM results \cite{RigidMultiblobs}. The components of $\sM{M}^{tt}$, $\sM{M}^{tr/rt}$, and $\sM{M}^{rr}$ are normalized by $1/\left(6\pi\eta R_h\right)$, $1/\left(6\pi\eta R_h^2\right)$, and $1/\left(8\pi\eta R_h^3\right)$, respectively. For problems involving rotation, we use the $m_M=m_D=6$ optimal ES monopole and dipole kernels (cf. Table~\ref{tab:optes}). If the problem only involves translation, we can use the optimal $m_M=4$ monopole kernel (cf. Table~\ref{tab:optes_trans}).

\begin{figure}[t]
  \centering
  \includegraphics{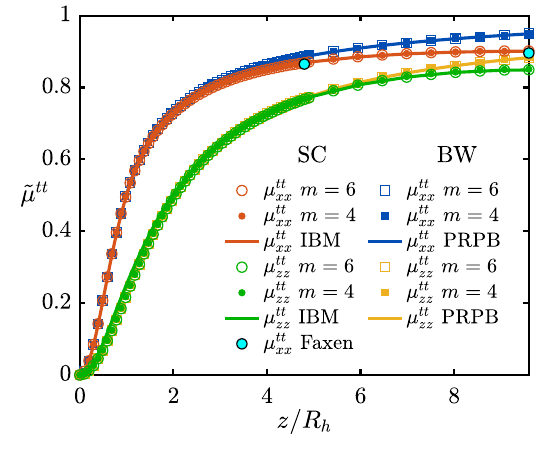}
  \caption{Normalized translation-translation self mobility, $\tilde{\mu}^{tt}=\left(6\pi\eta R_h\right)\mu^{tt}$, above the lower wall ($z\in[0,\tfrac{1}{2}H]$) in the bottom wall (BW) and slit channel (SC) geometries for $m=6$ (empty markers) and $m=4$ (filled markers). The reference data (shown with solid curves) for the BW and SC geometries are from a \marked{periodized Rotne-Prager-Blake (PRPB) tensor \cite{StokesianDynamics_Wall}} and numerical results from the immersed boundary method (IBM) \cite{RigidMultiblobs}, respectively. Fax\'en's results \marked{\cite{FaxenThesis}} for $\mu_{xx}^{tt}$ at $z=\tfrac{1}{4}H$ and $z=\tfrac{1}{2}H$ in the SC geometry are shown in cyan circles.}
  \label{fig:self_tt}
\end{figure}

\begin{figure}[t]
  \centering
  \includegraphics{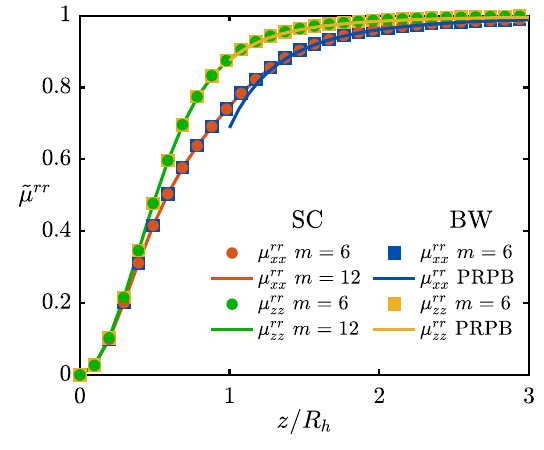}
  \caption{Normalized rotation-rotation self mobility, $\tilde{\mu}^{rr}=\left(8\pi\eta R_h^3\right)\mu^{rr}$, above the lower wall at $z=0$ in the BW and SC geometries. The reference results, from PRPB for the BW geometry and a twice resolved grid ($m=12$) for the SC geometry, are solid curves. Since $H\gg R_h$, the results are nearly identical between the BW and SC geometries.}
  \label{fig:self_rr}
\end{figure}

\subsection{Self mobility}\label{sec:valid:selfmob}
Here, we examine the behavior of the linear and rotational velocity of a particle given some prescribed forces or torques, for several particle positions relative to the wall(s), in both the BW and SC geometries.

First, we consider translation-translation self mobility. In the bottom wall (BW) geometry, we compare our numerical results to the periodized Rotne-Prager-Blake (periodized RPB) tensor. The RPB formulas \cite{StokesianDynamics_Wall} are for a half-space with no-slip conditions on the bottom wall, valid only for $z\ge R_h$. We periodize them by directly summing over $400$ images in the $x$ and $y$ directions to account for the periodic boundary conditions in our doubly-periodic simulations ($400^{2}$ images in total). We also compare to Fax\'en's power series expansions for the parallel self translational mobility of a sphere at half and quarter channel locations (see Eq.~(24) in \cite{RigidIBM}). In Fig.~\ref{fig:self_tt}, we plot the parallel (with respect to the wall(s)), $\mu_{xx}^{tt}=\mu_{yy}^{tt}$, and perpendicular, $\mu_{zz}^{tt}$, self mobilities of a blob. All of our numerical results agree with the corresponding references values for both $m=4$ and $m=6$ for $z\gtrsim 2R_h$. By the image construction \eqref{eq:kernelnearwall}, the particle mobility goes to zero smoothly as $z\to0$ for both the FCM and IBM, but in a kernel/solver dependent manner. Nevertheless, only small differences in $\mu_{zz}^{tt}$ near the wall are noticeable in the figure between the RPB, IBM, and FCM mobilities.

In Fig.~\ref{fig:self_rr} we show the parallel, $\mu_{xx}^{rr}=\mu_{yy}^{rr}$, and perpendicular, $\mu_{zz}^{rr}$, rotation-rotation mobility components. Again, our numerical results are in excellent agreement with the reference evaluations for $z\gtrsim 2R_h$. Since the wall corrections to the rotational components of the free space RPY tensor in the RPB tensor decay fast like $\left(R_h/z\right)^{3}$, $\mu_{xx}^{rr}$ and $\mu_{zz}^{rr}$ in the BW geometry are nearly indistinguishable from the same components in the SC geometry. As the blob approaches the wall, we see a noticeable difference between RPB and FCM for $\mu_{xx}^{rr}$. Lacking a reference result from other methods, in the SC geometry, we compare to a numerical reference computation on a doubly refined grid (i.e., $h\coloneqq\tfrac{1}{2}h$ and $m\coloneqq 2m$ such that the support $\alpha$ and the hydrodynamic radius remain constant).

\begin{figure}[t]
  \centering
  \includegraphics{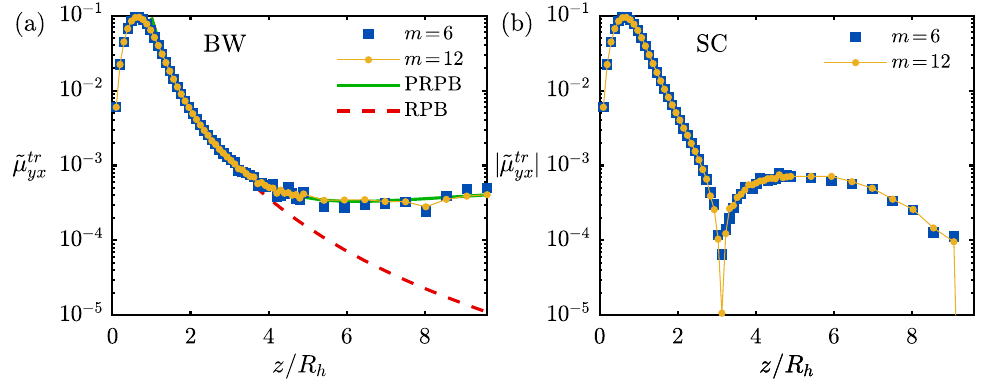}
  \caption{Normalized translation-rotation self mobility, $\tilde{\mu}_{yx}^{tr}=\left(6\pi\eta R_h^2\right)\mu_{yx}^{tr}$, above the lower wall ($z\in[0,\tfrac{1}{2}H]$) in the BW (a) and SC (b) geometries for $m=6$ and $m=12$ (a doubly-refined grid). For the BW geometry (a), we also show the results from PRPB and the half-space \emph{non-periodized} RPB tensors. Although not shown in this figure, for both BW and SC geometries, we find that the symmetric analog $\mu_{xy}^{rt}$ agrees with $\mu_{yx}^{tr}$ well within five digits. For the SC geometry (b), we show the absolute value due to a sign change for $z\gtrsim 3R_h$. The $x$ axis extends to $z=\tfrac{1}{2}H$, for which $\tilde{\mu}_{yx}^{tr}=0$ due to symmetry.}
  \label{fig:self_tr}
\end{figure}

In Fig.~\ref{fig:self_tr} we show $\mu_{yx}^{tr}$, the only nontrivial translation-rotation self coupling component. Interestingly, the mobility does not decay to zero as $z\to\infty$, as it would for a half-space domain; we display the non-periodized RPB kernel as dashed red lines in the left panel to highlight the effect of periodic boundary conditions in $x,y$. Furthermore, our numerical results for $m=6$, and to a lesser extent, even a doubly refined result with $m=12$, exhibit small numerical oscillations as the particle moves away from the wall.

To better understand the periodic artifacts in $\mu_{yx}^{tr}$, we analyze the contributions of periodic images of a particle to its self mobility. Consider the analytical form of the pairwise mobility $\nu_{yx}^{tr}$ (from the RPB tensor) for two particles, far away from each other, at the same height $z$,
\begin{equation}
  \label{eq:RPBanalytical}
\eta\nu_{yx}^{tr}(z)\approx\frac{1}{6\pi}\frac{3\hat{r}_z(\hat{r}_x)^2}{\norm{\V{r}}^2},
\end{equation}
where
\begin{equation}
  \V{r}=\begin{bmatrix}x^\jp{1}\!-\!x^\jp{2} & y^\jp{1}\!-\!y^\jp{2} & 2z\end{bmatrix}\tp,\quad \hat{\V{r}}=\frac{\V{r}}{\norm{\V{r}}}.
\end{equation}
We can approximate the sum of the RHS of \eqref{eq:RPBanalytical} over the periodic images in the $xy$ plane with an integral in polar coordinates. We find that as $z\to\infty$, the approximate sum converges to a constant $\sim L^{-2}$, and for $z\ll L$, it is $\sim z/L^3$. The self mobility $\mu_{yx}^{tr}$ itself decays like $R_h^2/z^4$. Therefore, the combined mobility (self plus images) is dominated by the self term for small $z$ and by the images for large $z$. This results in a minimum in $\mu_{yx}^{tr}(z)$, at roughly $z_{\text{min}}\sim\left(L^{3}R_{h}^{2}\right)^{1/5}$, followed by a plateau. If the particle goes above $z_{\text{min}}$, the translation-rotation self coupling is dominated by the periodic images.

For the slit channel, symmetry is always broken by the presence of the upper and lower walls, except at the exact center of the channel $z/R_h=H/2$. That is, $\mu_{yx}^{tr}(H/2)=0$, and we exclude this point from our plot due to the $\log$ scale on the $y$ axis. As expected, we have similar behavior between SC and BW near the wall, and the numerical oscillations are diminished for both the $m=6$ and $m=12$ reference computation in the SC geometry. Interestingly, there is a sign change in $\mu_{yx}^{tr}$ in the slit channel for $z\gtrsim 3R_h$ (likely due to periodic images), although the coupling is weak.

\subsection{Pair mobility}\label{sec:valid:pairmob}
Because we use a non-Gaussian tensor product kernel \eqref{eq:es}, the pair mobility for two blobs in an unbounded domain does not have a strictly isotropic form, even in the absence of discretization errors, and, furthermore, the pair mobility is difficult to compute analytically. In Appendix~\ref{sec:valid:pairmob:TP} we examine the pair mobility in the absence of walls numerically using a (large) TP domain. Our main conclusion is that the results obtained with the ES kernel are within a percent of those for a Gaussian kernel, and are isotropic to an accuracy of at least two digits for the kernel and grid choices we selected.  

\begin{figure}[t]
  \centering
  \setlength{\belowcaptionskip}{-10pt}
  \includegraphics{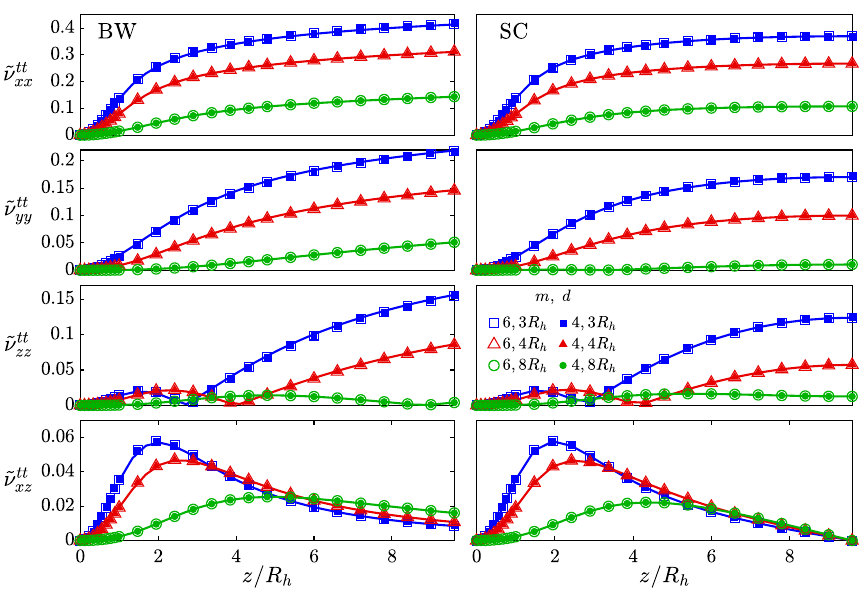}
  \caption{Normalized translation-translation pair mobility, $\tilde{\nu}^{tt}=\left(6\pi\eta R_h\right)\nu^{tt}$, in the BW (left panel) and SC (right panel) geometries, for a pair of particles, a distance $z$ from the bottom wall ($z\in[0,\tfrac{1}{2}H]$), the same $y$ coordinate, and distance $d=3R_h,4R_h,8R_h$ away from each other along the $x$ direction. Our results using the ES kernel are shown with empty and filled markers for $m=6$ and $m=4$, respectively. The reference results, PRPB for the BW geometry, and $m=12$ for the SC geometry, are solid curves.}
  \label{fig:pair_tt}
\end{figure}

Here we briefly investigate the behavior of the linear and rotational relative mobility of pairs of identical particles at the same distance $z$ from the bottom wall, and separated by a distance $d=3R_{h},4R_{h},8R_{h}$ in the $x$ direction. The presented results here serve solely as consistency checks and validation of our solver against available analytical and well-established numerical solutions. In Sec.~\ref{sec:colloidallayers:lanczos} we investigate in more detail some of the interesting behavior of the translation-translation pair mobility in a slit channel.

\begin{figure}[t]
  \centering
  \setlength{\belowcaptionskip}{-10pt}
  \includegraphics{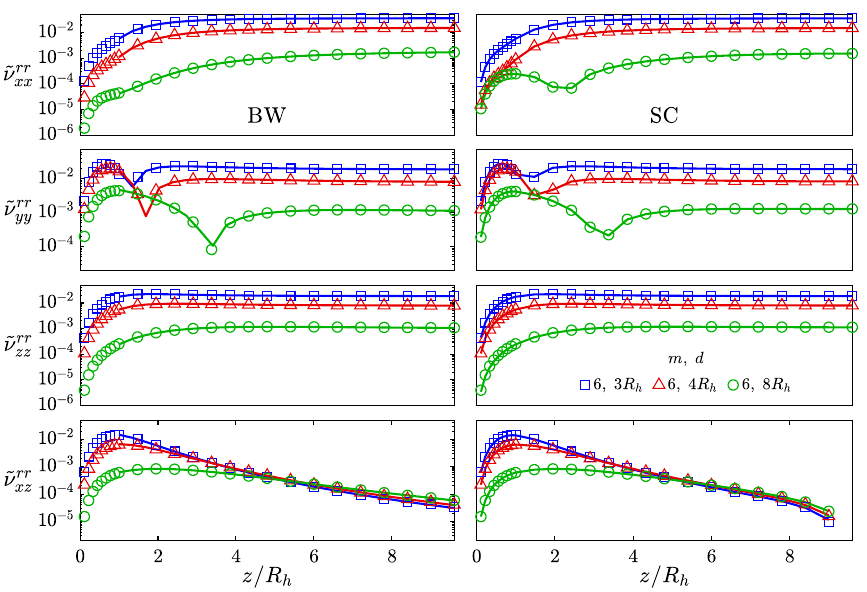}
  \caption{Normalized rotation-rotation pair mobility, $\tilde{\nu}^{rr}=\left(8\pi\eta R_h^3\right)\nu^{rr}$, for the same system as in Fig.~\ref{fig:pair_tt}. Our results using the ES kernel with $m=6$ are shown with empty markers. The reference results, PRPB for the BW geometry, and $m=12$ for the SC geometry, are solid curves. Note that $\mu_{xz}^{rr}$ goes to zero at the midplane due to symmetry.}
  \label{fig:pair_rr}
\end{figure}

\begin{figure}[t]
  \centering
  \setlength{\belowcaptionskip}{-10pt}
  \includegraphics{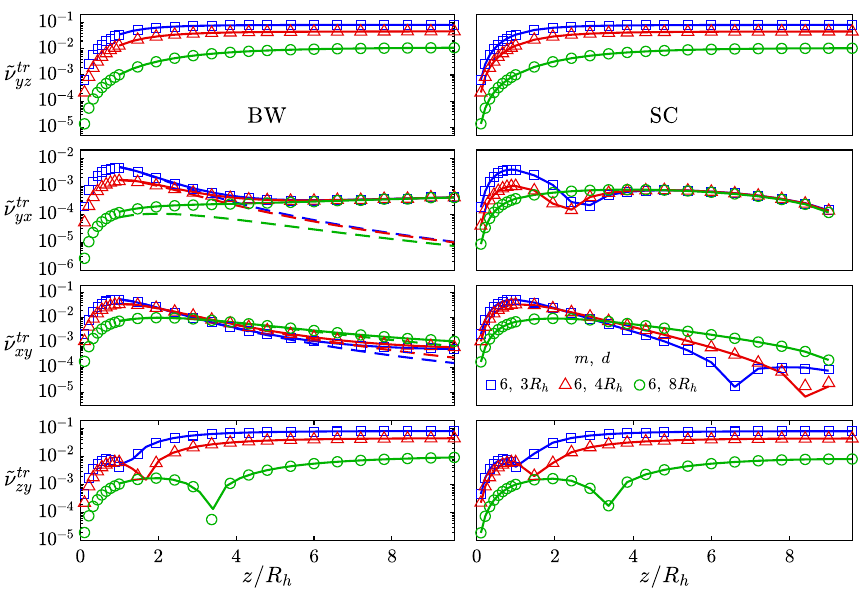}
  \caption{Normalized translation-rotation pair mobility, $\tilde{\nu}^{tr}=\left(6\pi\eta R_h^2\right)\nu^{tr}$, for the same system and labels as in Fig.~\ref{fig:pair_tt}. The half-space non-periodized RPB is also shown for mobility components $\nu_{yx}^{tr}$ and $\nu_{xy}^{tr}$ in BW as dashed curves. Although not shown here, for both BW and SC geometries, we find that the symmetric components of $\nu^{rt}$ agree with each other to well within five digits at each height. Note that the $yx$, $xy$, and $zy$ components of $\nu^{tr}$ go to zero at the midplane due to symmetry.}
  \label{fig:pair_tr}
\end{figure}

First, we examine the pair-wise components of $\nu^{tt}$ in Fig.~\ref{fig:pair_tt}. All numerical results compare well to the reference results, i.e., RPB in BW geometry, and a doubly-refined result in SC geometry, for both $m=4$ and $m=6$. For each mobility component considered, we see that when the particles are near the wall, greater separation leads to reduced magnitude of coupling. This is also the case further away from the bottom wall, except for $\nu_{xz}^{tt}$. In Fig.~\ref{fig:pair_rr}, we inspect the pair-wise components of $\nu^{rr}$. Our numerical results compare very well to the reference results in either geometry. As for translation-translation, the coupling decreases with particle separation, except in the case of the perpendicular-parallel component $\nu_{xz}^{rr}$. Lastly, in Fig.~\ref{fig:pair_tr}, we consider the pair-wise components of $\nu^{tr}$, and see that the reference results compare well to our numerical results. We highlight the same effects due to periodicity in $xy$ as seen in Fig.~\ref{fig:self_tr} through the non-periodized RPB kernel (dotted lines) for $\nu_{yx}^{tr}$ and $\nu_{xy}^{tr}$. Both $\nu_{yx}^{tr}$ and $\nu_{xy}^{tr}$ decay in a half-space as the particles are further from the bottom wall, but the periodicity in $xy$ leads to a plateau in the mobility.

\section{Colloidal Rollers}\label{sec:colloidallayers}

In this section, we consider a case study of a driven, dense suspension of colloidal microrollers as recently investigated by some of us \cite{RollersLubrication}. In this system, magnetic spherical colloidal particles sediment into a dense monolayer near a bottom wall/floor. A magnetic field is used to spin the particles with the axis of rotation parallel to the wall, which induces a translational motion of the particles along the wall due to the rotation-translation coupling induced by the presence of the floor (see $\mu_{yx}^{tr}$ in Fig.~\ref{fig:self_tr}, and $\nu_{yx}^{tr}$ and $\nu_{xy}^{tr}$ in Fig.~\ref{fig:pair_tr}). A collective driven steady state is established, and both experimental and numerical results show the existence of two layers of rollers, a slower layer close to the wall, and a faster layer above the slow one \cite{RollersLubrication}; see inset in Fig.~\ref{fig:micro_rollers}. Since the colloidal layer remains close to the wall in the $z$ direction but extends far in the $xy$ directions, this system is an ideal target for our doubly-periodic solver.

In \cite{RollersLubrication}, we developed a lubrication-corrected variant of Stokesian Dynamics method (without stresslet corrections) to simulate the collective driven dynamics of colloidal rollers, and found good agreements between computational and experimental results. In these prior studies, we computed the far-field hydrodynamic interactions using a GPU-accelerated direct pairwise summation of the Rotne-Prager-Blake tensor \cite{StokesianDynamics_Wall} generalized to account for torques/rotation. To approximate the doubly-periodic conditions, in these prior works \cite{MagneticRollers}, each particle interacted only with $3\times3=9$ periodic copies of every other particle. Because of the quadratic scaling of the direct summation, this increased the cost of the calculation by as much as two orders of magnitude (i.e., a factor of $9^2=81$), severely limiting scalability to larger numbers of particles.

Here, we apply our linearly-scaling method to the same problem. First, in Sec.~\ref{sec:colloidallayers:linear} we validate the linear scaling of our method up to the memory limitations on the GPU by periodically replicating a representative configuration from the steady state conditions found numerically in \cite{RollersLubrication}. In Sec.~\ref{sec:colloidallayers:lanczos} we study the convergence of an iterative method to compute the (far-field) Brownian (stochastic) particle displacements, since efficient and (log-)linear-scaling dynamic simulations require rapid convergence of the iterative method with a number of iterations independent of the number of particles. While we have previously established this for BW geometries \cite{MagneticRollers}, here we confirm this to be the case also in SC geometries, despite the slower decay of the hydrodynamic interactions with particle distance \cite{HI_Confined_Decay}. In Sec.~\ref{sec:colloidallayers:roller} we perform dynamic simulations with larger number of particles than previously feasible, and compare to previous results based on the Rotne-Prager-Blake tensor. We also add a top wall and investigate its influence on the collective dynamics and structure of the driven suspension.

\subsection{Linear scaling}\label{sec:colloidallayers:linear}
\begin{figure}[h]
  \centering
  \includegraphics{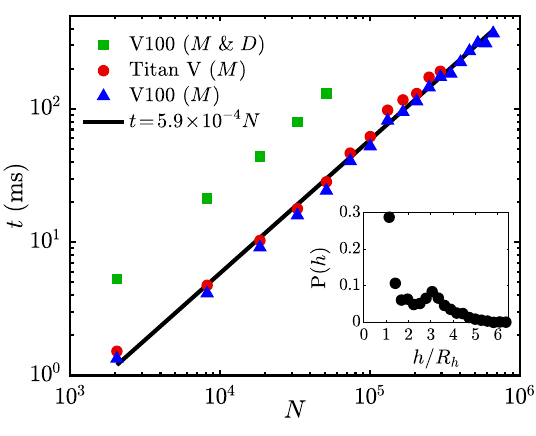}
  \caption{Time per mobility solve versus number of particles with parameters given in Table~\ref{tab:params-performance}, on V100 and Titan V NVIDIA GPUs. We replicate a system of $2048$ particles in the $xy$ plane. Particles are uniformly distributed in the planar direction, with an area packing fraction of $\phi=0.4$, and a bimodal (two-layer) distribution of heights $\mathrm{P}(h)$ above the wall, as shown in the inset. The biggest system contains $663552$ particles and takes $370\;\mathrm{ms}$ to compute with UAMMD on a V100 GPU.}
  \label{fig:replicas}
\end{figure}

In this section, we demonstrate the linear scaling of our GPU-based solver with the number of particles. Figure~\ref{fig:replicas} shows the computation time as we periodically replicate a representative configuration of $N=2048$ particles \cite{RollersLubrication} in the $xy$ planar direction to increase the number of particles. The original configuration before replication has a system size $L_{x/y}=L=128.8R_h$ and $H\approx 9R_h$ (see Table~\ref{tab:params-performance}). We test here the scaling for the case when there are only forces applied ($M$, monopole-only, kernel width $m=4$), and when both forces and torques are applied ($M\;\&\;D$, monopole and dipole terms, kernel width $m=6$). The computational cost for the dipole case is significantly larger than the monopole-only case, primarily because a larger grid size is required for kernel width $m=6$, and because additional FFTs are required in the present implementation of spreading and interpolation for the dipole terms \footnote{The alternative approach of differentiating the kernel is implemented in our CPU code.}. For the monopole-only case, the biggest system has $L=18\times128.8R_h=2318.4R_h$ ($H$ remains the same) and contains $663552$ particles, the maximum size that fits in the 32GB of memory in a V100 GPU. After $169$ replicas ($346112$ particles) the 12GB of memory available in the Titan V GPU are not enough.

\begin{table}[h]
\caption{Simulation parameters for the performance tests when only forces are applied (monopole-only, $M$), and when both forces and torques are applied (monopole and dipole, $M\;\&\;D$).}
\begin{center}
  \begin{tabular}{c?{\tabthickness}c|c|c|c|c|c|c|c}
               & $H$      & $L_{x/y}$  & $N_z$ & $N_{x/y}$ & $m_M$ & $\beta_M/m_M$ & $m_D$ & $\beta_D/m_D$\\
    \Xhline{\tabthickness}
    $M$        & $9.1R_h$ & $128.8R_h$ & $19$  & $150$     & $4$   & $1.87$        &       &\\
    $M\;\&\;D$ & $9.2R_h$ & $128.8R_h$ & $26$  & $216$     & $6$   & $1.38$        & $6$   & $2.3$
  \end{tabular}
\end{center}
\label{tab:params-performance}
\end{table}

Our performance test results clearly shows a linear scaling with the number of particles, allowing computations with on the order of a million particles in less than a second. This sort of large number of particles is infeasible with the direct summation method used by Sprinkle et al. \cite{RollersLubrication}. It is difficult to perform direct timing comparisons with the linear-scaling, FMM-based multicore, STKFMM code of Wen et al. \cite{STKFMM, FMM_wall}, since the performance of this CPU-based code depends heavily on the type of supercomputer and number of cores used, and optimization options. Nevertheless, rough timing comparisons suggest our simple GPU-friendly Stokes solver is at least an order of magnitude faster for this example on standard workstations.

\begin{figure}[t]
	\centering
	\resizebox{\textwidth}{!}{\includegraphics{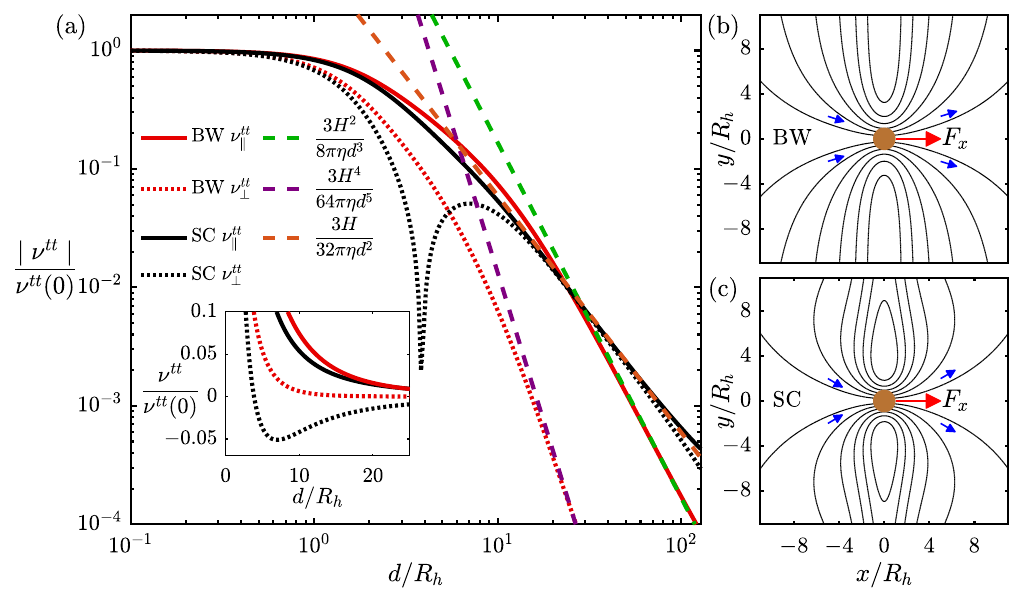}}
	\caption{(a) Two particles are placed at the same height $z=\tfrac{1}{2}H$, same $y$, and a distance $d$ apart from each other along the $x$ direction ($L=512R_h$, $H=8R_h$, $m=6$). One is pulled with a force in the $x$ or $y$ direction and the induced mobility on the other particle is measured in, respectively, $x$ (parallel component, $\nu_{\parallel}^{tt}=\nu_{xx}^{tt}$) and $y$ (perpendicular component, $\nu_{\perp}^{tt}=\nu_{yy}^{tt}$) directions. The dashed lines correspond to theoretical asymptotes for $d\gg R_h$. (b, c) Streamlines of a ghost particle velocity about a particle being pulled along the $x$ direction with a unit force $F_x$, calculated using the $x$ and $y$ velocity components $U_x^\jp{2}=\nu_{xx}^{tt}F_x$ and $U_y^\jp{2}=\nu_{yx}^{tt}F_x$.}
	\label{fig:pair_mobilities_decay}
\end{figure}

\subsection{Lanczos convergence}\label{sec:colloidallayers:lanczos}
Brownian Dynamics (BD) simulations require a method to generate Brownian particle ``velocities,'' which are Gaussian random variables with mean zero and covariance $\sim\sM{M}$, i.e., $\V{g}=\sM{M}^{\tfrac{1}{2}}\M{W}$, where $\sM{M}^{\tfrac{1}{2}}$ is a square root of the mobility matrix and $\M{W}$ is a collection of independent identically distributed (i.i.d) random numbers with zero mean and unit standard deviation \footnote{BD simulations in confined domains also require a method to generate a stochastic drift term $\sim\V{\partial}_{\V{r}}\cdot\sM{M}$, which can be done using two additional mobility solves per time step with Random Finite Differences \cite{MagneticRollers}.}. A Lanczos iterative algorithm \cite{SquareRootKrylov} can be used to estimate $\V{g}$ in the Krylov subspace of $\sM{M}$. Prior work has shown that the number of iterations required to achieve a certain tolerance is independent of the number of particles in suspensions of particles near a single wall (BW geometry, see Fig.~1 in \cite{MagneticRollers}), due to the hydrodynamic screening of the wall, but grows with the number of particles in the absence of a wall, i.e., in a TP environment (see Fig.~1 in \cite{SpectralRPY} and the inset of Fig.~1 in \cite{MagneticRollers}).

One could assume that the convergence rate of the Lanczos algorithm depends on how fast the hydrodynamic interactions decay with distance. In particular, we expect that the number of iterations is controlled by the slowest decaying hydrodynamic interactions. Therefore, for our problem of thin colloidal layers, we focus on the translation-translation components of the hydrodynamic interactions (no torques) in the $xy$ plane, for which the decay is slowest relative to the other components. In Fig.~\ref{fig:pair_mobilities_decay} we consider a pair of particles located at the same height $z=\tfrac{1}{2}H$, same $y$, and distance $d$ away from each other in the $x$ direction. We apply a unit force on one of the particles in either $x$ or $y$ directions and measure the resulting translational velocity of the other one in the same direction, i.e., $\nu_{xx}^{tt}$ and $\nu_{yy}^{tt}$, which we refer to as parallel, $\nu_{\parallel}^{tt}$, and perpendicular, $\nu_{\perp}^{tt}$, mobility components, respectively (see Fig.~\ref{fig:pair_mobilities_decay}(a)). We compare our results to the available asymptotic formulas in the literature for $d\gg R_h$. For SC geometry, both parallel and perpendicular mobilities approach the same asymptote $3H/\left(32\pi\eta d^2\right)$ (see Eq.~(51) in \cite{Liron1976}). For BW geometry, the parallel and perpendicular mobilities approach the asymptotes $3H^2/\left(8\pi\eta d^3\right)$ and $3H^4/\left(64\pi\eta d^5\right)$, respectively \cite{blake1971note}. Also, note that the data is normalized with the corresponding $\nu^{tt}(d=0)$, which is the mobility of a single particle located at height $z=\tfrac{1}{2}H$ \footnote{For large values of $H$ and in the absence of periodic effects, $\nu^{tt}(d=0)$ approaches the mobility of a particle in an unbounded domain, $1/\left(6\pi\eta R_h\right)$.}.

An interesting observation is that $\nu_{\perp}^{tt}$ becomes negative (presence of backflow) for $d/R_h\gtrsim 4$ in the SC geometry (see the inset of Fig.~\ref{fig:pair_mobilities_decay}(a)). To provide a better picture of this phenomena, in Figs.~\ref{fig:pair_mobilities_decay}(b) and (c) we show representative streamlines of a ghost particle velocity near a particle that is being dragged along the $x$ direction. We clearly see the existence of fluid vortices in the SC geometry case. Another important, and counter-intuitive, point is that the SC geometry shows a slower asymptotic decay rate compared to the BW geometry \cite{HI_Confined_Decay} ($d^{-2}$ versus $d^{-3}$). Such a slow decay warrants an examination of the Lanczos convergence rate in the SC geometry and how it compares with that in the BW and TP geometries.

\begin{figure}[t]
  \includegraphics{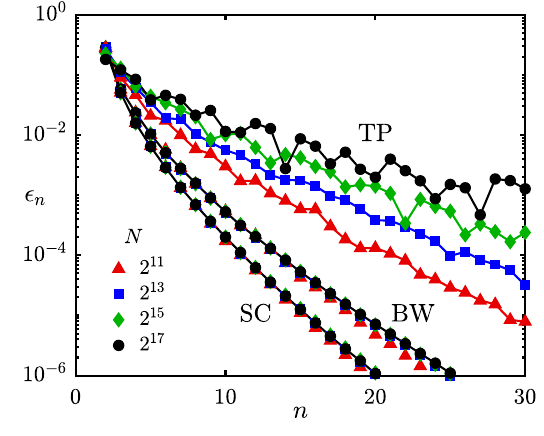}
  \caption{Relative error $\epsilon_n$ vs number of iterations $n$ of the Lanczos algorithm for a suspension of particles with the same configuration as in Fig.~\ref{fig:replicas}. Similar to Fig.~\ref{fig:replicas}, the system is replicated in the planar direction to study higher particle counts. The results for BW and SC geometries are obtained with the parameters given in Table~\ref{tab:params-performance} but $H=7.5R_h$. The TP result is for the same configuration but in triply-periodic mode, where we set the height of the (periodic) domain to $H=130R_h$.}
\label{fig:lanczos}
\end{figure}

Figure~\ref{fig:lanczos} shows the relative difference $\epsilon_n=\lVert\V{g}_n-\V{g}_{n-1}\rVert/\lVert\V{g}_{n-1}\rVert$ for the Lanczos algorithm versus the number of iterations $n$ in the BW, SC, and TP geometries. For the BW geometry, we observe the same phenomenon showcased in \cite{MagneticRollers}, where the hydrodynamic screening of the wall makes the convergence of the Lanczos algorithm essentially independent of the number of particles. The number of iterations required to achieve a $3$-digit accuracy remains less that $10$ for $N=2^{11}$ to $2^{17}$ particles. Surprisingly, the SC geometry shows a faster convergence rate than the BW one, despite the slower decay rate of the pair mobilities in this geometry (see Fig.~\ref{fig:pair_mobilities_decay}). Perhaps this can be explained by the fact that the slowly decaying component is \emph{not} of hydrodynamic origin, but rather comes from incompressibility; the actual hydrodynamic interactions are exponentially screened in the SC geometry \cite{HI_Confined_Decay}. Finally, consistent with prior work \cite{SpectralRPY}, the convergence rate drops significantly (and fails at times) for large TP systems due to the absence of the wall effects.

\begin{figure}[t]
  \includegraphics{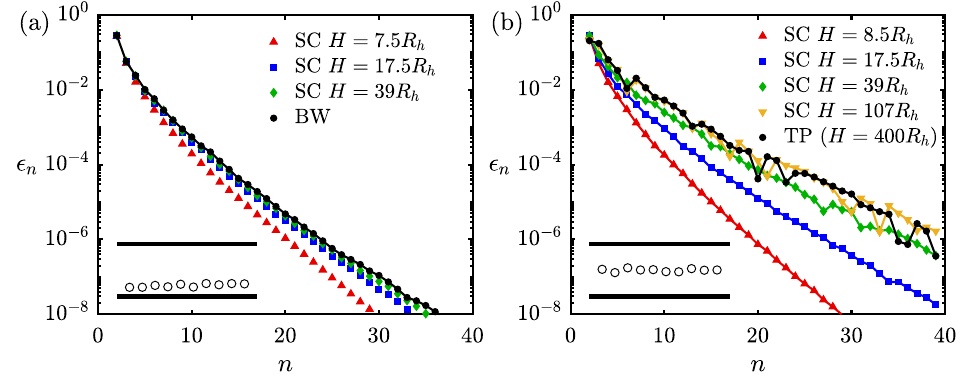}
  \caption{Lanczos convergence for a slit channel with increasing gap $H$, in two scenarios: (a) Particles ($N=2048$) are kept near the bottom wall and $H$ increases (which becomes increasingly similar to the bottom wall geometry); (b) Particles are kept at the middle of the channel and $H$ increases (which becomes increasingly similar to a TP geometry). Although not shown, above a relative error of $\sim10^{-6}$ the convergence is not affected by the floating point precision (single or double precision).}
\label{fig:lanczos_SC_to_BWTP}
\end{figure}

It should be noted that the convergence behavior for the SC geometry depends heavily on the thickness $H$ and the particle height $h$, as we explore Fig.~\ref{fig:lanczos_SC_to_BWTP}. We first place $2048$ particles near the bottom wall and increase the thickness $H$ between the two walls (see Fig.~\ref{fig:lanczos_SC_to_BWTP}(a)). We observe that as $H\to\infty$, the SC behavior approaches that of a BW geometry. We repeat this experiment with the particles placed at the midplane (see Fig.~\ref{fig:lanczos_SC_to_BWTP}(b)). This time, as $H\to\infty$, the system behavior approaches that of a TP system, as expected. Furthermore, we find that the convergence worsens as particles start to overlap the wall; this is a rare occurrence in BD simulations with steric repulsion.

\subsection{Microrollers}\label{sec:colloidallayers:roller}
In this section we use lubrication-corrected BD to study the dynamics of a dense suspension of colloidal microrollers in the BW and SC geometries.
\subsubsection{BW geometry}
As discussed above, in \cite{RollersLubrication} some of us described a method to apply near-field lubrication corrections (without stresslets) to a blob mobility matrix in the BW geometry, and how to efficiently use the corrected mobility matrix in the context of BD simulations. Blob-based descriptions of spherical particles provide efficient and adequate resolution of far-field hydrodynamic interactions at the expense of poor resolution for near-field hydrodynamics. Following Stokesian Dynamics, lubrication corrections are employed to fix the inaccurate near-field hydrodynamics and effectively capture the true hydrodynamics of nearly touching spheres, or a sphere near a bottom wall. Specifically, we can construct the lubrication corrected FCM mobility matrix as
\begin{equation}
\overline{\sM{M}}=\left(\R+\Delta\R\right)^{-1},\quad \Delta\R=\Rsup{A}-\Rsup{N}
\end{equation} 
where $\R=\sM{M}^{-1}$ is the resistance matrix computed using our FCM-based solver, and $\Rsup{A}$ and $\Rsup{N}$ are a superposition of pairwise resistance tensors for nearly touching surfaces (particle-particle or particle-wall) computed semi-analytically using the lubrication theory (or using a very refined blob-based method), and numerically by our FCM method \footnote{The method described in \cite{RollersLubrication} used the RPY tensor to compute $\R$ and $\Rsup{N}$.}, respectively. The subscript `$\sup$' signifies superposition of isolated pair interactions; each block $i,j$ of the resistance tensors $\Rsup{A}$ and $\Rsup{N}$ ($i,j\in\{1,\dots,N\}$) accounts for the near-field hydrodynamic interaction of nearby particles $i$ and $j$, $\Rpair{N/A}$, while the diagonal blocks correspond to the particle-wall interactions, $\Rbw{N/A}$.

To compute the correction $\Delta\R=\Rsup{A}-\Rsup{N}$, we need expressions for $\Rpair{N/A}$ and $\Rbw{N/A}$. Formulas for $\Rpair{A}$ and $\Rbw{A}$ are given in Appendix A of \cite{RollersLubrication}, while polynomial fits \footnote{Instead of polynomials fits, we opt to use linear interpolation on a precomputed set of values for each coefficient in both $\Rpair{N}$ and $\Rbw{N}$.} for the coefficients of $\Rpair{N}$ for FCM can be found in \cite{ForceCoupling_Bidisperse}. To compute $\Rbw{N}$, we simply use our FCM approach to generate a set a data for each coefficient vs height above the wall, taking care to use a large enough $L=L_{x/y}$ so that the periodic artifacts do not pollute the data, and linearly interpolate this data as needed.

\begin{figure}[t]
\centering
\includegraphics{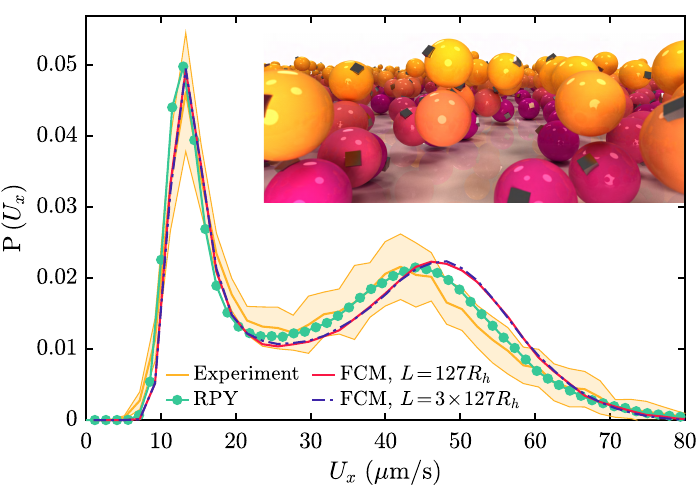}
\caption{A comparison of the experimentally measured (shown as a solid yellow line with a shaded area representing $95\%$ confidence bounds) and numerically computed velocity distributions in the direction of collective motion for a suspension of microrollers above a bottom wall. Simulated data are obtained from lubrication corrected BD in a doubly-periodic domain with $L=127 R_h$, using either the Rotne-Prager-Blake tensor, where periodicity is captured by directly including the $9$ nearest periodic images \cite{RollersLubrication}, our FCM approach, or our FCM approach in a $9\times$ larger domain. Inset: A typical configuration for a uniform suspension of microrollers at planar density $\phi=0.4$ and driving frequency $f=9\;\mathrm{Hz}$, suspended above a bottom wall. The hematite cube embedded in the particles is overemphasized for visual clarity and to show particle orientation. Low/slow particles are colored magenta while high/fast particles are colored yellow.} 
\label{fig:micro_rollers}
\end{figure}

To interrogate the effectiveness of using lubrication corrections with our FCM method,  we will study the uniform suspension of rotating particles above a bottom wall considered in Sec.~IV of \cite{RollersLubrication}. The ``microrollers'' are magnetic since they each have a cube of hematite embedded near their surface that exerts a torque when an external, rotating, magnetic field is applied to the whole suspension. The magnetic torque is strong enough that up to a certain frequency ($\sim 10\;\mathrm{Hz}$) the particles rotate with the same frequency of the applied field. We measure the particle velocity distribution $\mathrm{P}(U_x)$ in the positive $x$ direction in response to an applied magnetic field rotating in the $yz$ plane with a frequency of $9\;\mathrm{Hz}$, for a uniform suspension of microrollers with in plane packing fraction $\phi=0.4$. Figure~\ref{fig:micro_rollers} shows both the experimental measurements of $\mathrm{P}(U_x)$ and the numerical results using a lubrication corrected RPY mobility given in \cite{RollersLubrication}, along with results from a new simulation using our lubrication corrected FCM method. The new FCM method agrees reasonably well with both the experimentally measured velocity distribution as well as the old RPY results. However, there is a clear difference that stems from the fact that lubrication corrections are only approximate and have uncontrolled precision; perhaps including stresslets as well \cite{ForceCoupling_Lubrication,FluctuatingFCM_DC,SE_Multiblob_SD} would make the match better. Although it is tempting to compare to experimental data and choose one approximation over the other, it should be noted that some of the parameters in \cite{RollersLubrication} were chosen to match experimental data, while here we reuse the same parameters as in \cite{RollersLubrication} without re-estimating them.

\subsubsection{SC geometry}
In an SC geometry, the pair interactions remain the same, and one only needs to make modifications for the particle-wall interactions (compared to the BW geometry). In other words, we simply replace the near-field BW correction tensors $\dR{BW}=\Rbw{A}-\Rbw{N}$ with SC correction tensors, assuming grid invariance,
\begin{align}
\label{eq:dR_coeff}
&\dR{SC}=\Rsc{A}(h)-\Rsc{N}(h) = \nonumber\\
&\begin{bmatrix}
\Delta X^{tt}\left(h\right)\zhat \zhat^{T}+\Delta Y^{tt}\left(h\right)\left(\sM{I}-\zhat \zhat^{T} \right) & -\Delta Y^{tr}\left(h\right) \zhat\times \\
\Delta Y^{tr}\left(h\right) \zhat\times & \Delta X^{rr}\left(h\right)\zhat \zhat^{T}+\Delta Y^{rr}\left(h\right)\left(\sM{I}-\zhat \zhat^{T} \right)
\end{bmatrix},
\end{align}
where $h$ is the height of the particle's center above the bottom wall, $\zhat$ is the unit normal vector of the bottom wall, and $\zhat\times$ is the matrix that takes the cross-product with $\zhat$. However the available analytical formulas for the coefficients of $\Rsc{A}$ typically take the form of unwieldy and slowly convergent series \cite{RigidMultiblobs}, so we instead calculate the coefficients using a well-resolved multiblob representation of a sphere in a channel geometry. Recalling that only $\dR{SC}$ (but not $\Rsc{N/A}$ individually) is needed, all calculations should be done with the same unit cell in the $xy$ plane; this ensures that the results will rapidly converge in the limit as the periodic domain size $L_{x/y}=L\rightarrow \infty$. Note that the periodic domain size in all of these pre-calculations was taken to be $L=32R_h$ as this was determined to be sufficient to avoid any noticeable periodic artifacts.

\begin{figure}[t]
\centering
\includegraphics{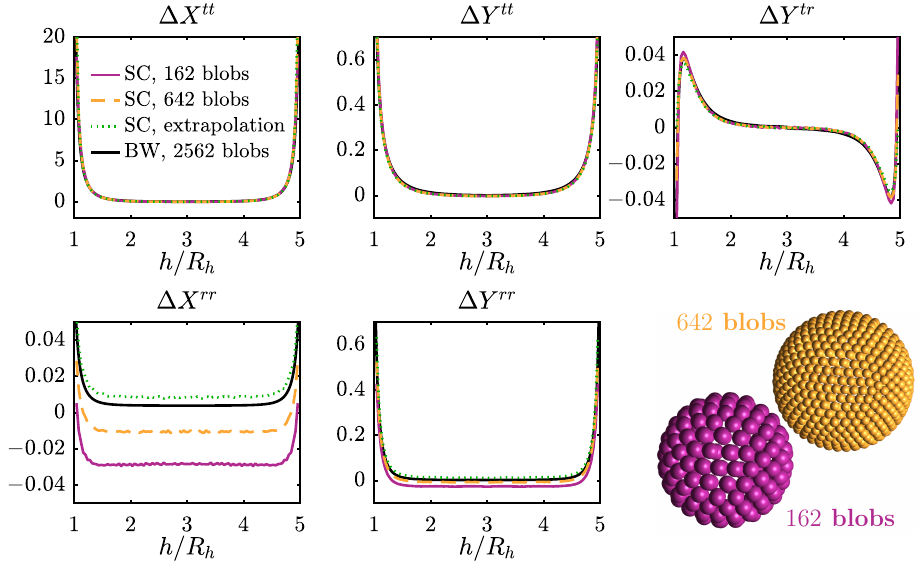}
\caption{For an SC geometry with height $H=6R_h$, each panel shows one of the five components in $\dR{SC}$ (see \eqref{eq:dR_coeff}) computed using our FCM-based solver and increasingly resolved rigid multiblob simulations \cite{RigidMultiblobs}. Richardson extrapolation is used to approximate more accurate values for $\dR{SC}$ than we can presently compute with the multiblob method, because large grid sizes are required. Each panel also shows components of the single wall corrections $\dR{BW}$ for the nearest wall, which provides a very reasonable approximation to the corrections in the SC geometry. Note that extrapolation is necessary (and not as accurate) for the components involving rotation since the rotational hydrodynamic radii of low-resolution multiblob spheres are different from their translational hydrodynamic radii \cite{RigidMultiblobs}, and computations with $2562$ blobs and $L=32R_h$ are infeasible due to limited GPU memory.} 
\label{fig:Rchan}
\end{figure}

Figure~\ref{fig:Rchan} shows each component of $\dR{SC}$ in \eqref{eq:dR_coeff} for a particle in an SC geometry with channel height $H=6R_h$ (notation for the coefficients follows \cite{SpheresStokes_JeffreyOnishi}). The components of $\dR{SC}$ are computed using an increasingly resolved multiblob \cite{RigidMultiblobs} representation of the sphere and extrapolated (Richardson) to infinite resolution. Figure~\ref{fig:Rchan} also shows each component of the single wall lubrication corrections to the nearest wall $\dR{BW}$, where $\Rbw{A}$ is computed with a very resolved $2562$ multiblob representation of a sphere \cite{RollersLubrication}. Note that it is also possible to add the lubrication corrections from both walls, but we have found this to be less accurate in thin channels because the corrections are doubled in the channel's center. Comparing the behavior of each component in $\dR{BW}$ with the behavior of $\dR{SC}$ shows that, at least for this representative problem with a channel width $H=6R_h$, superposing single wall corrections is a very reasonable approximation to the full slit channel corrections. Still, we opt to use the channel corrections in our simulations to emphasize that this approach can be adopted even for \emph{very} thin channels where superposing single wall corrections may provide a worse approximation. Numerically, we find that basing lubrication corrections on the nearest wall works surprisingly well even for channels as thin as $H=3R_h$.

\subsubsection{Effects of confinement on suspensions of microrollers}
A suspension of microrollers above a BW will self separate into a layer of slow moving particles that remain close to the wall, mainly due to gravity, and a fast layer of particles that are lifted above the slow layer, against gravity. These results are detailed in \cite{RollersLubrication} and reconfirmed in this section. Our spectral FCM approach now enables us to investigate the effect of further confinement on the velocity of a microroller suspension. In this section, we will use the same simulation parameters as those for the BW geometry considered in Sec.~\ref{sec:colloidallayers:roller}, but change the geometry to an SC with height $H=6R_h$. The $H=6R_h$ channel width was chosen based on the $98$th percentile of the particle height distribution obtained in the BW geometry (see inset in Fig.~\ref{fig:micro_rollers}), so that the top wall only provides a minor steric constraint on the suspension. 

\begin{figure}[t]
\centering
\includegraphics{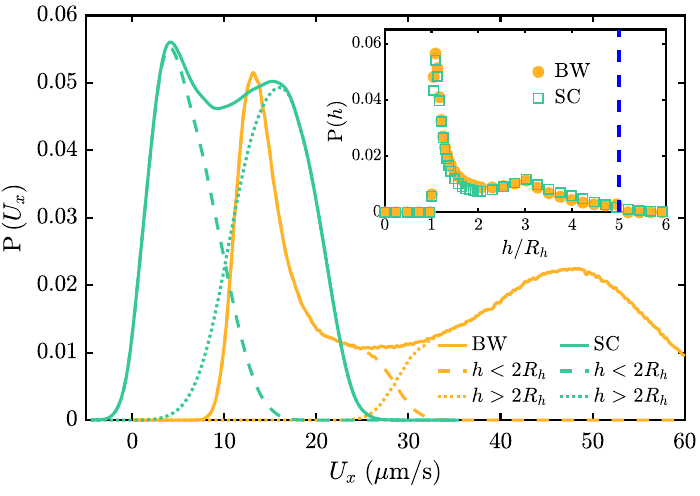}
\caption{Distributions of particle height (inset) and velocity (main figure) for a microroller suspension above a single wall (BW, gold line with circular markers), and a slit channel with width $H=6R_h$ (SC, teal line with square markers). Particle velocity distributions are shown for each channel geometry where each distribution is also broken into sub-distributions for low (close to the bottom wall, $h<2R_h$) and high ($h>2R_h$) particles. These sub-distributions show that regardless of domain geometry the `high' particles make up the mode of `fast' particles in the suspension (where `fast' is to be interpenetrated as having a large, \emph{positive} $U_x$). Inset: Dimensionless particle height ($h/R_h$) distributions are shown for each channel geometry. The dashed line demarcates maximum possible particle height in the SC geometry.} 
\label{fig:ChanDists}
\end{figure}

Figure~\ref{fig:ChanDists} compares the velocity distributions ($\mathrm{P}(U_x)$) of the driven particles in a suspension confined by a single BW, or an SC with height $H=6R_h$. The velocity distributions in each case are also broken up into sub-distributions consisting of particles whose height above the BW is above/bellow a certain threshold (specifically, $h \lessgtr 2R_h$). This shows that the faster particles in the suspension (those which have the largest \emph{positive} $U_x$) are also the highest particles in the suspension. The inset of Fig.~\ref{fig:ChanDists} indicates that the particle height distributions in the BW and SC ($H=6R_h$) geometries are fairly similar, while the main figure shows large differences in the velocity distributions for these two geometries. Although both distributions are bimodal, there are no really fast particles in the SC case. Coincidentally, the ``fast'' layer is at a distance of about $3R_h$ from the bottom wall. If it were not for the ``slow'' layer of particles breaking the vertical symmetry, particles at $h=3R_h$ would not move when spinning. We can thus conclude that this significant slowing of the suspension velocity in the SC geometry compared with the BW geometry is due to the presence of a top wall, which changes the character of the flow generated by driving the suspension.

\section{Conclusions}
In this paper we have developed a variant of the force coupling method (FCM) for computing hydrodynamic interactions among blobs/particles suited for doubly-periodic geometries including ones that have an open aperiodic direction. We employed a more compact non-Gaussian kernel to reduce the grid size required to resolve the particles. Our novel spectral Stokes solver uses FFTs in the two periodic directions and Chebyshev transforms in the aperiodic direction, and is suited for efficient implementation on GPUs using a single 3D FFT for each Fourier-Chebyshev transformation. We used our method to show that a dense monolayer of colloidal microrollers splits in two layers even in a slit channel, but moves substantially slower.

The key ideas used in our Stokes solver, namely, the use of the Dirichlet-to-Neumann map to obtain boundary conditions in the aperiodic direction, as well as the use of correction solves to enforce boundary conditions, can be used to generalize the solver to other boundary conditions in the aperiodic direction, such as partial slip (e.g., superhydrophobic surfaces) or free slip boundaries (e.g., a fluid-fluid interface), including cases when there are particles on both sides of the boundary. However, it is important to point out that the method relies in a key way on the double periodicity in two directions to obtain a system of uncoupled one dimensional boundary value problems, and is therefore tailored to doubly-periodic geometries.

\marked{It is possible to extend our work to account for the contribution of stresslets, as already implemented in the traditional FCM method; see Eqs.~(7) and (11) in \cite{ForceCoupling_Stokes}. While we have not done so, there is no difference as far as our method or any FCM method is concerned between torque and stresslet; they are spread with the same kernel, and both are spread and interpolated using derivatives of the kernel. The real challenge is how to efficiently solve for the stresslets; see the work of Fiore and Swan \cite{SE_Multiblob_SD} for some ideas toward effective preconditioning. Additionally, one might include fluctuations (Brownian motion) with stresslets as described in Appendix B of \cite{BrownianBlobs}, as well as in \cite{FluctuatingFCM_DC} and \cite{SpectralSD}.}

Two key challenges for future work are to combine the force coupling method with Ewald splitting, and to devise a method to generate stochastic displacements for BD without iterations. This would allow to decouple the blob hydrodynamic radius from the grid size, and make stochastic simulations of comparable cost to deterministic simulations. Both of these challenges have been tackled for the RPY kernel in the Positively Split Ewald method \cite{SpectralRPY} for triply periodic domains. Both challenges have also partially been addressed in a low-order immersed boundary approach \cite{DISCOS_Periodic,DISCOS_Confined} including for slit channel geometries, but the accuracy is difficult to control and the Ewald splitting is approximated by a particle-particle-particle-mesh approach. 

Considerable progress has been made on handling the first challenge due to its importance in boundary integral methods for Stokes flow. Spectral Ewald Stokes solvers for the singular Stokeslet and Rotlet have been developed extensively by the group of Tornberg \cite{SpectralEwald_Stokes,SpectralEwald_Wall}, including recent work that unifies arbitrary periodicities in different directions \cite{SE_MixedPeriodicityStokes}. One can directly apply these methods to the bottom wall geometry we studied here using a method of images, but the finite size of the particles and the potential overlap of the particles with each other and the wall need to be accounted for either using the RPY kernel (as in recent fast multipole methods \cite{FMM_wall}) or an FCM-like approach. To our knowledge, an Ewald splitting for FCM has not yet been developed even for triply periodic domains. It seems quite plausible that by combining ideas from this paper (and our work on Poisson solvers \cite{PoissonDP}) with extensive work in the field of FCM and Spectral Ewald methods, in the future the community can develop an Ewald splitting method for finite-size blobs capable of handling all doubly-periodic geometries.

The second challenge, generating the action of the square root of the mobility explicitly without costly iterations, has hardly been addressed for non triply periodic domains. We have been able to achieve this for doubly periodic unconfined geometries using the approach developed in this work, however, it required switching to a Galerkin formulation of the boundary value problems in the unbounded direction. This not only leads to dense linear systems, instead of the pentadiagonal ones in the method presented here, but it also couples all three velocities and pressure in one linear solve for each planar Fourier mode. This makes the approach less suited to memory-limited GPUs and more costly if sufficient memory is available. Furthermore, accounting for the wall(s) remains a nontrivial challenge for future work.

\section*{Acknowledgments}
We thank Leslie Greengard, Kyongmin Yeo, and Amanda Howard for helpful discussions. This work was supported by the NSF under award DMS-2011544 and through a Research and Training Group in Modeling and Simulation under award RTG/DMS-1646339.

\newpage
\appendix

\setcounter{equation}{0}
\renewcommand{\theequation}{A\arabic{equation}}
\section{Slit channel geometry with slip velocity}\label{sec:appendixA-sc}
For the SC geometry, for increased generality and because of its relevance to microfluidics, we show how to impose partial slip boundary conditions at $z=0$ and $z=H$:
\begin{subequations}
\label{eq:bcslip}
\begin{align}
  \V{u}|_{z=0}-\V{u}_0^\W & =\ell_0\begin{bmatrix}\displaystyle{\frac{\partial u(x,y,0)}{\partial z}} & \displaystyle{\frac{\partial v(x,y,0)}{\partial z}} & 0\end{bmatrix}\tp,\label{eq:bcslip0}\\
  \V{u}|_{z=H}-\V{u}_H^\W & =-\ell_H\begin{bmatrix}\displaystyle{\frac{\partial u(x,y,H)}{\partial z}} & \displaystyle{\frac{\partial v(x,y,H)}{\partial z}} & 0\end{bmatrix}\tp.\label{eq:bcslipH}
\end{align}
\end{subequations}
Here, $\V{u}_{0/H}^\W=\begin{bmatrix}u_{0/H}^\W & v_{0/H}^\W & 0\end{bmatrix}\tp$ are the wall velocities, and $\ell_{0/H}$ are the slip lengths, with $\ell=0$ denoting a no-slip wall.

For simplicity, we write the equations explicitly only for $u$, with identical equations for $v$ with the swap $x\leftrightarrow y$. Note that the DP solver remains the same as that of the single wall geometry, noting that particle kernels have images in both walls. It is the correction solve that takes care of the boundary conditions.

\subsection{Correction Solve}\label{sec:correctionsolve-sc}
We have $\V{u}=\V{u}_\DP+\V{u}_\corr$, which upon substitution into \eqref{eq:bcslip}, and some rearrangement, yields
\begin{subequations}
  \label{eq:bcslip-corr}
  \begin{align}
    &\V{u}_\corr|_{z=0}-\ell_0\begin{bmatrix}\displaystyle{\frac{\partial u_\corr(x,y,0)}{\partial z}} & \displaystyle{\frac{\partial v_\corr(x,y,0)}{\partial z}} & 0\end{bmatrix}\tp =\begin{bmatrix}u_0(x,y) & v_0(x,y) & w_0(x,y)\end{bmatrix}\tp,\label{eq:bcslip0-corr}\\
    &\V{u}_\corr|_{z=H}+\ell_H\begin{bmatrix}\displaystyle{\frac{\partial u_\corr(x,y,H)}{\partial z}} & \displaystyle{\frac{\partial v_\corr(x,y,H)}{\partial z}} & 0\end{bmatrix}\tp =\begin{bmatrix}u_H(x,y) & v_H(x,y) & w_H(x,y)\end{bmatrix}\tp,\label{eq:bcslip0-corr}
  \end{align}
\end{subequations}
where,
\begin{subequations}
  \label{eq:u0uHformula}
  \begin{align}
    u_0(x,y)&=u_0^\W-u_\DP(x,y,0)+\ell_0\frac{\partial u_\DP(x,y,0)}{\partial z},\\
    w_0(x,y)&=-w_\DP(x,y,0),\\
    u_H(x,y)&=u_H^\W-u_\DP(x,y,H)-\ell_H\frac{\partial u_\DP(x,y,H)}{\partial z},\\
    w_H(x,y)&=-w_\DP(x,y,H).
  \end{align}
\end{subequations}

Following a similar method as the one described in Sec.~\ref{sec:correctionsolve}, one can find the general solutions to $\hat{p}_\corr$ and $\hat{\V{u}}_\corr$, the pressure and velocity in Fourier space, as
\begin{align}
  \hat{p}_\corr(\V{k},z)&=C_0e^{-kz}+D_0e^{kz},\label{eq:psolfs-corr-sc}\\
  \hat{u}_\corr(\V{k},z)&=-\frac{C_0\i k_x}{2\eta k}ze^{-kz}+\frac{D_0\i k_x}{2\eta k}ze^{kz}+C_xe^{-kz}+D_xe^{kz},\label{eq:usolfs-corr-sc}\\
  \hat{w}_\corr(\V{k},z)&=\frac{C_0}{2\eta}ze^{-kz}+\frac{D_0}{2\eta}ze^{kz}+C_ze^{-kz}+D_ze^{kz}.\label{eq:wsolfs-corr-sc}
\end{align}

To find the constants, we start by writing the boundary conditions \eqref{eq:bcslip-corr} in Fourier space:
\begin{subequations}
  \label{eq:bcslip-corr-fs}
  \begin{align}
    \hat{u}_\corr(\V{k},0)-\ell_0\frac{\partial\hat{u}_\corr(\V{k},0)}{\partial z}=\hat{u}_0(\V{k}),\\
    \hat{w}_\corr(\V{k},0)=\hat{w}_0(\V{k}),\\
    \hat{u}_\corr(\V{k},H)+\ell_H\frac{\partial\hat{u}_\corr(\V{k},H)}{\partial z}=\hat{u}_H(\V{k}),\\
    \hat{w}_\corr(\V{k},H)=\hat{w}_H(\V{k}).
  \end{align}
\end{subequations}
Substituting \eqref{eq:usolfs-corr-sc} and \eqref{eq:wsolfs-corr-sc} into \eqref{eq:bcslip-corr-fs} results in the following equations for the unknown constants
\begin{subequations}
  \label{eq:6eqs-for-constants}
  \begin{align}
    \left(\frac{\i k_x \ell_0}{2\eta k}\right)C_0-\left(\frac{\i k_x \ell_0}{2\eta k}\right)D_0+\left(1+k\ell_0\right)C_x+\left(1-k\ell_0\right)D_x&=\hat{u}_0(\V{k}),\\
    C_z+D_z&=\hat{w}_0(\V{k}),\\
    \frac{\i k_xHe^{-kH}}{2\eta k}\left(k\ell_H-\frac{\ell_H}{H}-1\right)C_0+\frac{\i k_xHe^{kH}}{2\eta k}\left(k\ell_H+\frac{\ell_H}{H}+1\right)D_0&\nonumber\\
    +e^{-kH}\left(1-k\ell_H\right)C_x+e^{kH}\left(1+k\ell_H\right)D_x&=\hat{u}_H(\V{k}),\\
    \frac{He^{-kH}}{2\eta}C_0+\frac{He^{kH}}{2\eta}D_0+e^{-kH}C_z+e^{kH}D_z&=\hat{w}_H(\V{k}).
  \end{align}
\end{subequations}

Furthermore, the continuity equation in Fourier space \eqref{eq:continuityfs-corr} can be written at $z=0,H$:
\begin{equation}
  \i k_x\hat{u}_\corr(\V{k},0/H)+\i k_y\hat{v}_\corr(\V{k},0/H)+\frac{\partial\hat{w}_\corr(\V{k},0/H)}{\partial z}=0.
  \label{eq:continuityfs-corr-sc-0H}
\end{equation}
Inserting the general solutions \eqref{eq:usolfs-corr-sc}--\eqref{eq:wsolfs-corr-sc} into the above equations yields two more equations for the unknown coefficients:
\begin{subequations}
  \label{eq:2eqs-for-constants}
  \begin{align}
    \mkern-30mu\frac{C_0}{2\eta}+\frac{D_0}{2\eta}+\i k_xC_x+\i k_xD_x+\i k_yC_y+\i k_yD_y-kC_z+kD_z&=0,\\
    \mkern-30mu\frac{e^{-kH}}{2\eta}C_0\!+\!\frac{e^{kH}}{2\eta}D_0\!+\!\i k_xe^{-kH}C_x\!+\!\i k_xe^{kH}D_x\!+\!\i k_ye^{-kH}C_y\!+\!\i k_ye^{kH}D_y\!-\!ke^{-kH}C_z\!+\!ke^{kH}D_z&=0.
  \end{align}
\end{subequations}

Now we have a system of eight equations given by \eqref{eq:6eqs-for-constants} (with the corresponding equations for $v$) and \eqref{eq:2eqs-for-constants} which can solved to find the eight unknown constants $C_0,D_0,C_{x/y/z},D_{x/y/z}$ for each $\V{k}=\begin{bmatrix}k_x & k_y\end{bmatrix}\tp\neq\V{0}$.

\subsection{Zero mode ($\V{k}=\V{0}$)}\label{sec:k0mode-sc}
The zero mode $\V{k}=\V{0}$ solutions to the velocity component normal to the walls, $\hat{w}(\V{0},z)$, and the pressure, $\hat{p}(\V{0},z)$, remain the same as those of the single wall geometry, given by \eqref{eq:psolfs-k0} and \eqref{eq:wsolfs-k0}. However, the $\V{k}=\V{0}$ mode of the tangential velocity components will be different. Fourier transform of the slip boundary conditions \eqref{eq:bcslip} is
\begin{subequations}
\label{eq:bcslipfs}
\begin{align}
  \hat{\V{u}}(\V{k},0)-\hat{\V{u}}_0^\W(\V{k}) & =\ell_0\begin{bmatrix}\displaystyle{\frac{\partial \hat{u}(\V{k},0)}{\partial z}} & \displaystyle{\frac{\partial \hat{v}(\V{k},0)}{\partial z}} & 0\end{bmatrix}\tp,\label{eq:bcslip0fs}\\
  \hat{\V{u}}(\V{k},H)-\hat{\V{u}}_H^\W(\V{k}) & =-\ell_H\begin{bmatrix}\displaystyle{\frac{\partial \hat{u}(\V{k},H)}{\partial z}} & \displaystyle{\frac{\partial\hat{v}(\V{k},H)}{\partial z}} & 0\end{bmatrix}\tp.\label{eq:bcslipHfs}
\end{align}
\end{subequations}
Therefore, similar to the single-wall case (cf. \eqref{eq:momentumfs-k0-u} and \eqref{eq:bvp-u-k0-bcs}), one can find $\hat{u}(\V{0},z)$ (and analogously $\hat{v}(\V{0},z)$) via solving the following BVPs:
\begin{equation}
  \mkern-10mu\frac{\partial^2\hat{u}(\V{0},z)}{{\partial z}^2}=-\frac{\hat{f}(\V{0},z)}{\eta},\quad\hat{u}(\V{0},0)-\ell_0\frac{\partial\hat{u}(\V{0},0)}{\partial z}=\hat{u}_0^\W(\V{0}),\quad\hat{u}(\V{0},H)+\ell_H\frac{\partial\hat{u}(\V{0},H)}{\partial z}=\hat{u}_H^\W(\V{0}).\label{eq:bvp-uv-k0-sc}
\end{equation}

\setcounter{equation}{0}
\renewcommand{\theequation}{B\arabic{equation}}
\section{Translation-translation pair coupling}\label{sec:valid:pairmob:TP}
In this Appendix, we numerically compute and investigate the invariance under rotation and translation of the pair translational mobility, $\sM{M}^{tt}$, for the ES kernel in a large triply periodic domain. We will examine our numerical approximation to $\sM{M}^{tt}$ as a function of $d=\norm{\V{d}}=\lVert\V{d}^\jp{1}-\V{d}^\jp{2}\rVert$, where $\V{d}^\jp{1}$ and $\V{d}^\jp{2}$ are the positions of the two particles. If the kernel is rotationally invariant (which our tensor product ES kernel \eqref{eq:es} is not), then $\sM{M}^{tt}$ would have the invariant form
\begin{equation}
\sM{M}^{tt}(\V{d})=f(d)\sM{I}+g(d)\frac{\V{d}\otimes\V{d}}{d^2},\label{eq:frgr}
\end{equation}
for some functions $f(d)$ and $g(d)$. Lomholt and Maxey \cite{ForceCoupling_Stokes} derived analytical forms for \emph{free space} $f(d)$ and $g(d)$ when particles are represented by Gaussian force envelopes (in the absence of discretization artifacts),
\begin{subequations}
  \label{eq:frgrgauss}
\begin{align}
f(d) & = \frac{1}{8\pi\eta d}\left(\left(1+2\frac{R_h^2}{\pi d^2}\right)\mathrm{erf}\left(\frac{d\sqrt{\pi}}{2R_h}\right)-2\frac{R_h}{\pi d}\exp\left(-\frac{\pi d^2}{4R_h^2}\right)\right)\label{eq:frgauss}\\
g(d) & = \frac{1}{8\pi\eta d}\left(\left(1-6\frac{R_h^2}{\pi d^2}\right)\mathrm{erf}\left(\frac{d\sqrt{\pi}}{2R_h}\right)+6\frac{R_h}{\pi d}\exp\left(-\frac{\pi d^2}{4R_h^2}\right)\right),\label{eq:grgauss}
\end{align}
\end{subequations}

The invariant form \eqref{eq:frgr} does not strictly apply in practice for our FCM method for several reasons. The first is that the tensor product kernel we use is not rotationally invariant as a Gaussian kernel is. The second is that rotational invariance is broken for a finite system such as a periodic box. Lastly, there are discretization artifacts due to the finite grid resolution in the Stokes solver, especially for the small kernel widths like $m=4$. Here we numerically demonstrate that our method produces a pair mobility that matches \eqref{eq:frgrgauss} to at least two digits.

\begin{figure}[t]
  \centering
  \includegraphics{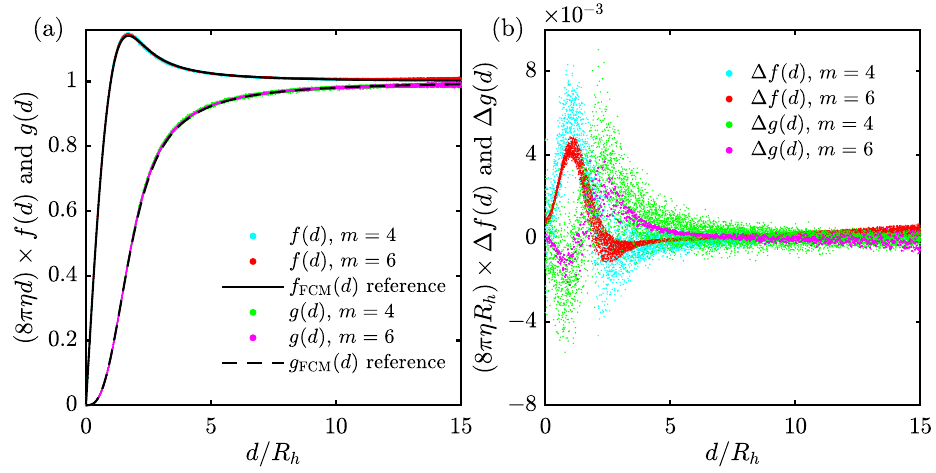}
  \caption{(a) Numerically calculated, normalized, mobility functions $\left(8\pi\eta d\right) f(d)$ and $\left(8\pi\eta d\right) g(d)$, for two spheres in a large triply-periodic domain with ES kernel, versus the distance $d$ between the two spheres, along with reference FCM curves given by \eqref{eq:frgrgauss} for the continuum model in free space with Gaussian envelopes. (b) Normalized difference between the numerical and analytical values of $f(d)$ and $g(d)$ (i.e., $\left(8\pi\eta R_h\right)\left[f(d)-f_{\mathrm{FCM}}(d)\right]$ and $\left(8\pi\eta R_h\right)\left[g(d)-g_{\mathrm{FCM}}(d)\right]$, respectively) versus $d$.}
  \label{fig:frgr_numeric}
\end{figure}

The numerical domain is a cubic box of side length $L=200$ and uniform grid spacing $h=1$. We consider particles represented by the optimal $m=6$ monopole kernel with $R_h=1.55$, or the optimal $m=4$ monopole kernel with $R_h=1.205$ (see Table~\ref{tab:optes_trans}). The two particles always use the same kernel types. We generate $10^{4}$ random particle pair configurations a distance $d\in[0,25]$ from each other. For each pair, we compute the pair parallel and perpendicular mobilities $\nu_{\parallel}$ and $\nu_{\perp}$ with respect to the direction $\V{p}$. According to \eqref{eq:frgr}, we have
\begin{align}
  \nu_{\parallel} &\approx f(d;L)+g(d),\\
  \nu_{\perp} &\approx f(d;L),
\end{align}
where $f(d;L)$ depends on $L$ due to the finite size of the triply-periodic box. Following Hasimoto's corrections for the self mobility in triply periodic domains \cite{Mobility2D_Hasimoto} (see \eqref{eq:tp-velocity-linear-angular}), we correct the numerically calculated $f(d;L)$ for $d\ll L$ by
\begin{equation}
  \label{eq:fdcorrect}
  f(d)\approx f(d;L)+\frac{1}{6\pi\eta R_h}\left(2.84\frac{R_h}{L}\right).
\end{equation}

In Fig.~\ref{fig:frgr_numeric}, we compare the corrected $f(d)$ and $g(d)$ for the ES kernel to the free space $f(d)$ and $g(d)$ for Gaussian force envelopes given by \eqref{eq:frgrgauss}. We see a good match for both $m=4$ and $m=6$ monopole ES kernels (see Fig.~\ref{fig:frgr_numeric}(a)). Figure~\ref{fig:frgr_numeric}(b) shows the normalized difference between the numerically calculated $f(d)$ and $g(d)$ by the ES kernel and the traditional FCM mobility functions \eqref{eq:frgrgauss}. We see that the data clusters around a curve that is systematically different from \eqref{eq:frgrgauss}, but the difference is on the order of a percent or less. Also, the scatter of $f(d)$ and $g(d)$ due to imperfect translational and rotational invariance are well controlled within two digits. Note that the increased error at high $d/R_h$ is due to periodic boundary conditions artifacts.



%

\end{document}